\documentclass[review]{elsarticle}

\usepackage{lineno,hyperref}

\usepackage[utf8]{inputenc}
\usepackage{multicol}
\usepackage{tabularx}
\usepackage{graphicx}
\usepackage{float}
\usepackage{caption}
\usepackage{subcaption}
\usepackage{multirow}
\usepackage{textcomp}
\usepackage{xcolor} 
\usepackage[normalem]{ulem} 
\usepackage{verbatim} 
\usepackage{amsmath} 
\usepackage{geometry}
\begin{document}
\newcommand{\R}[2][red]{\textcolor{#1}{#2}}
\begin{frontmatter}

\title{Feature Analysis for Machine Learning-based IoT Intrusion Detection}


\author[mymainaddress]{Mohanad Sarhan\corref{mycorrespondingauthor}}
\cortext[mycorrespondingauthor]{Corresponding author}
\ead{m.sarhan@uq.net.au}

\author[mymainaddress]{Siamak Layeghy}
\ead{siamak.layeghy@uq.net.au}

\author[mymainaddress]{Marius Portmann}
\ead{marius@itee.uq.edu.au}

\address[mymainaddress]{The University of Queensland, St Lucia QLD 4072}

\begin{abstract}

Internet of Things (IoT) networks have become an increasingly attractive target of cyberattacks. Powerful Machine Learning (ML) models have recently been adopted to implement network intrusion detection systems to protect IoT networks. For the successful training of such ML models, selecting the right data features is crucial, maximising the detection accuracy and computational efficiency. This paper comprehensively analyses feature sets' importance and predictive power for detecting network attacks. Three feature selection algorithms: chi-square, information gain and correlation, have been utilised to identify and rank data features. The attributes are fed into two ML classifiers: deep feed-forward and random forest, to measure their attack detection performance. The experimental evaluation considered three datasets: UNSW-NB15, CSE-CIC-IDS2018, and ToN-IoT in their proprietary flow format. In addition, the respective variants in NetFlow format were also considered, i.e., NF-UNSW-NB15, NF-CSE-CIC-IDS2018, and NF-ToN-IoT. The experimental evaluation explored the marginal benefit of adding individual features. Our results show that the accuracy initially increases rapidly with adding features but converges quickly to the maximum. This demonstrates a significant potential to reduce the computational and storage cost of intrusion detection systems while maintaining near-optimal detection accuracy. This has particular relevance in IoT systems, with typically limited computational and storage resources.

\end{abstract}


\end{frontmatter}

\begin{multicols}{2}

\section{Introduction}

Internet of Things (IoT) is the application of interconnected digital objects such as sensors and processing units \cite{boyes2018industrial}. It enables organisations to operate efficiently and reliably using advanced digital technologies such as machine-to-machine communication and Artificial Intelligence (AI). IoT integrates information technology with operational technology to provide industries with automation and optimisation capabilities. IoT aims to improve information systems communication quality, productivity, and efficiency. The deployed IoT endpoints collect real-time information to help organisations in their decision-making processes, as machines are often equipped with AI tools to perform and automate operational tasks efficiently. Although adopting the IoT can empower organisations, having robust defensive security strategies in place is a current research issue \cite{yu2019survey}. IoT ecosystems attract new challenges due to integrating the physical and digital worlds, where the cost of error can be devastating, such as operational disruption and financial loss. Many deployed endpoints provide an increased attack surface area for malicious actors to access operating systems. A notorious attack example is the Ukrainian power grid in December 2015 \cite{case2016analysis}, where attackers gained remote access to the grid control unit and disrupted the power of more than 230,000 users. Another example is the deployment of the WannaCry ransomware into the IoT network of a Taiwanese chip manufacturer in 2018 \cite{alladi2020industrial} that resulted in damages of approximately \$170 million. Therefore, organisations are looking for improved methods to protect their IoT networks with an increasing dependence on digitalisation and automation.

Network Intrusion Detection Systems (NIDSs) are implemented in IoT networks to preserve the three security principles of information systems; confidentiality, integrity, and availability \cite{4656556}. They scan and analyse network traffic to detect signs and patterns that indicate a potential attack. Traditional NIDSs aim to match incoming traffic signatures with a predetermined list of known malicious signatures. This usually leads to a high Detection Rate (DR) against known attacks; however, these systems have been proven to be unreliable against previously unseen attacks (zero-day attacks) or new variants of existing attacks \cite{amoli2016unsupervised}. Machine Learning (ML) has gained increasing attention over the past few years, mainly due to the increased capabilities of ML algorithms and data availability. ML-based NIDSs represent an emerging technology to detect malicious network traffic as they penetrate the target network and have been successfully adopted by researchers. The development of such advanced detection capabilities achieves a promising attack detection accuracy \cite{garcia2009anomaly}. In the design phase of such systems, the learning models aim to extract and learn complex security events from network data samples. The data samples are presented in network data features such as the number of in/out bytes and in/out packets.

Feature selection is critical in building an efficient ML-based network attack detection model. A smaller set of features can result in more efficient collection and storage, which is necessary for a high-speed IoT network environment \cite{zhang2013effective}. In other words, the performance of ML-based NIDSs depends on the quality and integrity of the data used to train and evaluate ML models. This paper aims to assess the feature importance of six publicly available NIDS datasets; UNSW-NB15, NF-UNSW-NB15, ToN-IoT, NF-ToN-IoT, CSE-CIC-IDS2018, and NF-CSE-CIC-IDS2018. Three feature selection algorithms, namely Chi-square (CHI), Information Gain (IG), and Correlation (COR), have been used to identify the top features in each dataset. The features have been fed into two ML classifiers: Deep Feed Forward (DFF) and Random Forest (RF), to evaluate their attack detection performance. The rest of the paper is organised as follows. Related works are discussed in the following section, and the importance of feature analysis in NIDS datasets is explained in Section \ref{bfs}. Section \ref{em} explains the comprehensive methodological approach adopted in our paper. Finally, the results are represented and discussed in Section \ref{r}. The paper's essential contribution is the comprehensive analysis of feature importance across six NIDS datasets and two ML-based classifiers. The main findings are that a) a smaller feature subset than the currently used in state-of-the-art works can be sufficient to achieve the same attack detection performance, b) there is a high variability of feature ranking across feature selection algorithms and NIDS datasets, and c) certain features have an unrealistically high predictive power and should not be used in the ML models' training to avoid experimental bias and achieve reliable detection results.

\section{Related Work}
\label{rw}

This section discusses key related works that have aimed to identify the optimal network traffic feature sets in NIDS datasets. Zhang et al.  \cite{zhang_wu_gao_wang_xu_liu_2018} applied a Deep Feed Forward  (DFF) neural network on the UNSW-NB15 dataset after utilising a denoising AutoEncoder (AE) to identify the ten most representative features. The overall classification accuracy achieved is 98.80\%, with a False Alarm Rate (FAR) of 0.57\%. However, a relatively low DR of 94.43\% and the lack of evaluation using other ML classifiers limits the effectiveness of the proposed optimal feature set. 

In \cite{bagui_kalaimannan_bagui_nandi_pinto_2019}, the authors explored the effect of performing k-means clustering and adding correlation-based feature selection methods to Naive Bayes (NB) and Decision Tree (DT) classifiers. The aim was to find the optimal subset of features for the experiment. The results showed that the correlation-based feature selection method improved the results of the NB model but had no effect on the DT classifier. The study also found which features contribute the most to detecting each attack type present in the UNSW-NB15 dataset. 

In \cite{aljawarneh_aldwairi_yassein_2018}, the authors measured the IG of each feature in the NSL-KDD dataset. The authors select eight features based on an IG threshold value of 0.40 to potentially increase the accuracy of their ML model. Their proposed hybrid model uses J48, Meta Pagging, RF, REPTree, AdaBoostM1, DecisionStump and NB classifiers. The accuracy achieved by binary classification is 99.81\%. The authors also applied the wrapper method \cite{chandrashekar2014survey} to select the best feature subset, and the classification accuracy obtained is 98.56\% in the multi-class classification case. 

In \cite{moustafa_slay_2015}, the authors used Association Rule Mining (ARM) as a feature selection algorithm to generate the most relevant list of features in the UNSW-NB15 and KDD9 datasets. The ARM method is implemented using the \textit{apriori} algorithm defined by a set of 100 rules to select 11 features for each attack class. The NB and Expectation-Maximisation (EM) models perform the classification to evaluate the selected features. The study shows that ARM performed well on the KDD99 dataset. However, the classifiers cannot detect some attack types from the UNSW-NB15 data set.

In \cite{binbusayyis_vaiyapuri_2019}, an ensemble feature selection approach has been designed that applies four techniques, i.e., correlation, consistency, IG, and Tanimoto distance \cite{bolon2019ensembles}. The generated features of four datasets are combined using a subset combination method, depending on how often the feature selection methods identify the feature. The authors confirmed the reliability of the selected features by analysing the variance (ANOVA) method. The combination of features identified three or more times achieved the best metrics across the datasets. 

A-Zewairi et al.  \cite{al-zewairi_almajali_awajan_2017} applied a DFF on the UNSW-NB15 dataset to measure the effectiveness of the Gedeon method \cite{gedeon1997data} in selecting the essential features, which are used in the experiment and listed in the paper, and grouped according to their level of importance. The authors also compared three activation functions (ReLU, tanh, and max out), each with and without a dropout mode, showing that the ReLU function without a dropout mode achieved the best results. The best accuracy of 98.99\% and a FAR of 0.56\% is obtained by training the model using the top 20\% of the essential features.

Mogal et al. \cite{mogal_ghungrad_bhusare_2017} applied NB and Logistic Regression (LR) classifiers to the UNSW-NB15 and KDDcup99 datasets, choosing accuracy and prediction time as the defining metrics. They used two feature selection techniques: Central Point and ARM apriori algorithms, to select the most highly ranked features. However, the paper does not mention the number of features chosen to train or test the models. Although these feature selection and reduction techniques did not increase the level of accuracy of the classifiers, they notably decreased their prediction times. The best levels of accuracy of 99.94\% and 99.96\% were obtained using LR on the reduced KDDcup99 dataset and NB on the complete feature set of the UNSW-NB15 dataset, respectively.

In \cite{9096621}, the authors designed a bidirectional long and short-term memory with a multi-feature layer (B-MLSTM) model to secure IoT networks from low-frequency and multistage attacks. The proposed system learns the pattern of the attack interval through historical data, using sequence and stage feature layers to effectively detect attacks at different intervals. A double-layer reverse unit is introduced to update the detection model's parameters. The proposed scheme has a lower false alarm rate than related works on three IoT datasets.

In general, extensive research has been done to identify the optimal feature sets in different NIDS datasets. However, the experimental methodology of the papers mentioned above is limited and requires a presumption of several selection criteria, such as a particular feature selection technique and rules (threshold) of selection.
The necessity of deciding a predetermined selection method assuming that it would generalise across datasets, limits its reliability due to the statistical and feature differences of NIDS datasets \cite{sarhan2020netflow}.

In contrast to previous work, this paper presents a more systematic and extensive evaluation of NIDS feature sets, considering a more significant number of feature selection algorithms, feature subsets, and NIDS datasets. Table \ref{rwt} highlights this by showing the number of ML classifiers, the number of feature selection algorithms, the number of feature sets and the number of NIDS datasets considered in this paper compared to key-related works. This paper considers recently released NIDS datasets (ToN-IoT, NF-UNSW-NB15, NF-ToN-IoT and NF-CSE-CIC-IDS2018), for which feature set analysis has yet to be published.

\end{multicols}
\begin{table}[ht]
\scriptsize
\centering
\caption{Related Papers}
\label{rwt}
\begin{tabular}{|l|l|l|l|l|}
\hline
\multicolumn{1}{|c|}{\textbf{Paper}} &
  \multicolumn{1}{c|}{\textbf{\begin{tabular}[c]{@{}c@{}}No. of ML \\ Models\end{tabular}}} &
  \multicolumn{1}{c|}{\textbf{\begin{tabular}[c]{@{}c@{}}No. of Feature \\ Selection Algorithms\end{tabular}}} &
  \textbf{\begin{tabular}[c]{@{}l@{}}No. of Considered \\ Feature Subsets\end{tabular}} &
  \textbf{\begin{tabular}[c]{@{}l@{}}No. of NIDS \\ Datasets\end{tabular}} \\ \hline
Zhang et al. \cite{zhang_wu_gao_wang_xu_liu_2018}            & 1          & 1          & 1           & 1          \\ \hline
Bagui el al. \cite{bagui_kalaimannan_bagui_nandi_pinto_2019} & 2          & 1          & 1           & 1          \\ \hline
Aljawarneg et al. \cite{aljawarneh_aldwairi_yassein_2018}    & 7          & 2          & 1           & 1          \\ \hline
Moustafa et al. \cite{moustafa_slay_2015}                    & 2          & 1          & 1           & 2          \\ \hline
Binbusayyis et al. \cite{binbusayyis_vaiyapuri_2019}         & 1          & 4          & 1           & 4          \\ \hline
Mogal et al. \cite{mogal_ghungrad_bhusare_2017}              & 2          & 2          & 1           & 2          \\ \hline
Xinghua et al. \cite {9096621}                      & 1          & N/A          & N/A           & 3          \\ \hline
\textbf{This paper}                         & \textbf{2} & \textbf{3} &  \textbf{8-15} & \textbf{6} \\ \hline
\end{tabular}%
\end{table}

\begin{multicols}{2}

\section{Datasets}
\label{bfs}

An ML-based NIDS learns and extracts the patterns of benign and attack flows in network traffic. These patterns are learned in terms of data features extracted from network traffic. The performance of an ML-based NIDS is determined by choice of network traffic features in the design process. Choosing a better feature set leads to a higher-performing ML model and better traffic classification performance \cite{fahad2013toward}. In IoT networks, devices deployed at the edge might not be equipped with sufficient computational power and storage capacity to extract many network traffic features. Therefore, removing redundant and irrelevant features can reduce the computational and storage demand of NIDSs in scenarios with resource-constrained edge devices, such as IoT. A careful choice of feature subsets can minimise the classification performance loss compared to the point where the complete set is used. As this paper demonstrates, in some cases, an optimally chosen feature subset can provide even better classification performance than the entire feature set.

Due to privacy and practical concerns in obtaining real-world labelled NIDS datasets, researchers have generated many synthetic benchmark NIDS datasets on testbed networks. These datasets have been widely used to evaluate ML-based network traffic classifiers and intrusion detection systems. The current NIDS datasets contain features that provide relevant predictive power to classify network traffic. The feature sets provided in the most relevant NIDS datasets are quite different due to the other choices of feature extraction tools and design decisions made by the creators of these datasets. This paper adopts a comprehensive approach to identify the importance of features for the considered NIDS datasets. Unlike other research papers, our approach is not limited to any presumptions regarding the optimal feature set, as multiple feature selection algorithms, ML classifiers, and many feature subsets are explored. This paper considers six publicly available and recently published NIDS datasets for our experiments.

%

\begin{itemize}

\item \textbf{ToN-IoT} \cite{fesz-dm97-19} - This is a heterogeneous dataset released by the Cyber Range Lab of the Australian Centre for Cyber Security (ACCS) in 2019. It is a comprehensive dataset that includes the telemetry data of an IoT network. Several portions of the dataset contain different traces of IoT services, network traffic, and Operating System (OS) logs. The data was generated using a realistic network testbed. The dataset contains several attack scenarios such as backdoor, DoS, Distributed DoS (DDoS), injection, Man In The Middle (MITM), password, ransomware, scanning, and Cross-Site Scripting (XSS). Bro-IDS tool, now called Zeek, was used to extract 44 network traffic features. 

\item \textbf{NF-ToN-IoT} \cite{sarhan2020netflow} - This dataset is a variant of the ToN-IoT dataset, but with a NetFlow feature set. To create the NF-ToN-IoT dataset, the packet capture file (pcap) of the network part of the original ToN-IoT dataset was converted to NetFlow format using the nprobe tool \cite{Ntopng2017}. This dataset contains the same attack types as the original file but with a NetFlow-based feature set with eight features.

\item \textbf{UNSW-NB15} \cite{moustafa-slay-2015} - This is a very widely used dataset in the NIDS research community, released in 2015 by the ACCS Cyber Range Lab. The authors utilised the IXIA PerfectStorm tool to generate benign traffic. The Argus and Bro-IDS tools were used to extract 49 network traffic features. Nine attack scenarios are implemented in a testbed: fuzzers, analysis, backdoor, Denial of Service (DoS), exploits, generic, reconnaissance, shellcode, and worms. 

\item \textbf{NF-UNSW-NB15} \cite{sarhan2020netflow} - As in the case of NF-ToN-IoT, the original packet captures (pcaps) of the UNSW-NB15 dataset were converted to NetFlow format using the nProbe tool to generate a NetFlow-based dataset named NF-UNSW-NB15. The dataset includes eight Netflow features and nine different attack scenarios, as in the original UNSW-NB15 dataset.

\item \textbf{CSE-CIC-IDS2018} \cite{sharafaldin-habibi-lashkari-ghorbani-2018} - This is an NIDS dataset released by a project involving the Communications Security Establishment (CSE) \& Canadian Institute for Cybersecurity (CIC) in 2018. The testbed is configured to replicate a realistic organisational network comprising five departments and a server room comprising various application servers. Different attack types, such as brute-force, bot, DoS, DDoS, infiltration, and web attacks, are conducted on an external network to simulate realistic external attack scenarios. The CICFlowMeter-v3 tool was implemented to extract 77 features to create the dataset. 

\item \textbf{NF-CSE-CIC-IDS2018} \cite{sarhan2020netflow} - As in the above two NetFlow-based datasets, the pcap files of the CSE-CIC-IDS2018 dataset were converted to the NF-CSE-CIC-IDS2018 dataset with eight NetFlow-based features. The dataset includes the same 14 attack types as the original CSE-CIC-IDS2018 dataset.

\end{itemize}


\section{Experimental Methodology}
\label{em}
This paper considers six publicly available synthetic datasets, each reflecting modern attack types and benign network traffic. Three widely used and highly effective feature selection algorithms: chi-square, IG, and correlation, are applied to the complete set of features. The features are then ranked based on importance and identified by each feature selection method. Finally, two well-known ML classifiers: Deep Feed Forward (DFF) neural network and Random Forest (RF), are used to measure the attack detection performance of the different feature subsets. The experiment starts with considering the single most highly ranked feature, the top two most highly ranked features sub-set, up to the set with the 15 top-ranked features.

The flow identifier features, such as flow IDs, timestamps, source/destination IP addresses, and port numbers, were dropped to avoid bias towards attacking/victim nodes and applications. Moreover, any non-numerical features were converted to numerical values using label encoding. Min-Max normalisation was applied to scale the feature values from 0 to 1. The datasets have been divided into 70\%-30\% for training and testing purposes. Finally, stratified 5-fold cross-validation is used to achieve reliable results. The DFF classifier was implemented using the TensorFlow library. In combination with the Scikit-learn library, Python programming language was used to implement the feature selection methods and the RF classifier.

\subsection{Feature Selection}
Feature selection reduces the number of features in a dataset without losing essential or relevant information. Supervised feature selection algorithms aim to rank the features by measuring their statistical relationship to the class label \cite{DASH1997131}. The features with the highest ranking hold most of the valuable information needed to predict the type of traffic class. Feature extraction, also known as dimensionality reduction techniques such as PCA and LDA have not been considered. This is due to the fact that they project a full feature set into a smaller set in a different feature space, rather than identifying the important features in the original feature space. Therefore, the following three supervised feature selection algorithms have been used to rank the features in the considered NIDS datasets.

\begin{itemize}
    \item \textbf{Chi-square} \cite{sharpe2015chi} - The chi-square $\chi_{c}^{2}$ test is used to measure the independence of a feature and its respective class label. The chi-square measures how the expected label $E$ and the feature $O$ deviate from each other. The degree of freedom $c$ determines whether the null hypothesis can be rejected. A high chi-square value indicates that the hypothesis of independence should be rejected, as the feature and class depend on it, and that the feature should be used in the classification experiments. Equation \ref{c3} shows the relevant definition.
    
%
%
\begin{equation}
\chi_{c}^{2}=\sum \frac{\left(O_{i}-E_{i}\right)^{2}}{E_{i}}
\label{c3}
\end{equation}

\item \textbf{Information Gain} \cite{mackay2003information} - A parametric formula is used to measure the mutual dependence between the features and the label; higher values indicate a higher contribution in making the correct classification and hence a valuable feature. Equation \ref{c2} defines the IG, or mutual information $I(X ; Y)$, for labels $X$ and features $Y$, where $H(X)$ is the entropy for $X$ (i.e., uncertainty about $X$),  and $H(X \mid Y)$ is the conditional entropy for $X$ given $Y$. IG is a symmetric measure of mutual dependence between two feature sets and measures the reduction in the uncertainty (entropy) about the traffic labels if we are given a set of features. This reduction in uncertainty (entropy) corresponds to a gain in information, as shown in Equation~\ref{c2}.
\begin{equation}
I(X ; Y) =H(X)-H(X \mid Y)
\label{c2}
\end{equation}

\item \textbf{Correlation} \cite{hall1999correlation} - In this feature selection or ranking method, Pearson's correlation coefficient algorithm is used to compute the linear correlation score between the labels and features. 
It is defined as the covariance of the labels ($X$) and features ($Y$), divided by the product of their standard deviations, as shown in Equation \ref{c}.
\begin{equation}
\rho_{X, Y}=\frac{{cov}(X, Y)}{\sigma_{X} \sigma_{Y}}
\label{c}
\end{equation}

\end{itemize}

\subsection{Machine Learning}
The building of ML models following a supervised classification method includes two processes; training and testing. During the training phase, the model is trained with labelled network features to extract and learn patterns. The testing phase involves evaluating the detection reliability of the model by measuring its performance in classifying unseen traffic into an attack or benign class. The predictions are compared with the actual labels to evaluate the ML model using the standard evaluation metrics listed and defined in Table \ref{metrics}, in which TP, FP, TN, and FN represent the number of True Positives, False Positives, True Negatives and False Negatives respectively. For our experiments, we considered the following two classifiers due to their wide usage and effective performance in the detection of network intrusions:

\begin{itemize}
    \item \textbf{Deep Feed Forward (DFF)} - A DFF neural network model consisting of 3 hidden layers is used as a network traffic classifier in our experiments. The input layer is set to match the input's number of dimensions, and each hidden layer contains ten nodes using the ReLU activation function. The output layer is a single-node sigmoid classifier. Between each layer, the model implements a dropout rate of 20\% to avoid overfitting. Finally, the binary cross-entropy loss function was used in conjunction with the Adam optimiser. 
    
    \item \textbf{Random Forest (RF)} - An RF classifier is used, consisting of 50 randomised  Decision Tree classifiers on various subsamples of the dataset. The Gini impurity function measures a split's quality and configures the Decision Trees. 

\end{itemize}

\end{multicols}


\begin{table}[ht]\scriptsize
\caption{Evaluation metrics}
\centering
\begin{tabular}
{|c|c|c|} 
\hline
\textbf{Metric} & \textbf{Equation} & \textbf{Explanation} \\
\hline 
Recall & 
\parbox{3cm}{
\begin{equation*}
    \frac{TP}{TP+FN}
    \label{eq:DR}
\end{equation*} 
}
& Fraction of attacks detected = \textbf{DR} 

\\ [0.25cm]
\hline
Precision &
\parbox{3cm}{
\begin{equation*}
    \frac{TP}{TP+FP}
    \label{eq:Precision}
\end{equation*}
}& 
The fraction of detected attacks to total alarms
\\
\hline
F1-score &
\parbox{3.5cm}{
\begin{equation*}
    2 \times \frac{Recall\times Precision}{Recall + Precision}
    \label{eq:F1-score}
\end{equation*}
} &
The weighted average of the precision and recall \\
\hline
Accuracy &
\parbox{3.5cm}{
\begin{equation*}
     \frac{TP+TN}{TP+FP+TN+FN}
    \label{eq:Accuracy}
\end{equation*}
}
& Fraction of the correctly classified samples \\
\hline
False Alarm Rate &
\parbox{3cm}{
\begin{equation*}
    \frac{FP}{FP+TN}
    \label{eq:FAR}
\end{equation*}
} & Fraction of benign samples classified as attack \\
\hline

{Prediction Time} &
\parbox{3cm}{
\begin{equation*}
    \frac{\text{Total Prediction Time}}{\text{Number of Sample}}
    \label{eq:time}
\end{equation*}
} & Number of µs to predict a single sample \\
\hline

Area Under the Curve (AUC) & \multicolumn{2}{c|}{Area under the Receiver Operating Characteristics (ROC) curve \cite{narkhede2018understanding}}
\\
\hline

\end{tabular}
\label{metrics}
\end{table}



\begin{multicols}{2}

\section{Results}
\label{r}

In our experiments, all features of each NIDS dataset are ranked based on their importance using the three feature selection algorithms, i.e., chi-square, IG, and correlation. The classification performance using only the most highly ranked feature is initially considered. We then continue to add individual features to the feature subset in rank order. All feature subsets from size 1 to 15 are evaluated, and the corresponding classification performance for our DFF and RF classifiers is measured.

\end{multicols}

\subsection{UNSW-NB15 and NF-UNSW-NB15}
\subsubsection{Full Feature Set}
\label{uns}

\begin{table}[ht]\scriptsize
\centering
\caption{UNSW-NB15 ranked features}
\label{ft}
\begin{tabular}{l|llll|}
\cline{2-5}
 &
  \multicolumn{1}{l|}{\textbf{Rank}} &
  \multicolumn{1}{l|}{\textbf{Chi-square}} &
  \multicolumn{1}{l|}{\textbf{Information Gain}} &
  \textbf{Correlation} \\ \hline
\multicolumn{1}{|l|}{\textbf{\begin{tabular}[c]{@{}l@{}}UNSW-\\ NB15\end{tabular}}} &
  \begin{tabular}[c]{@{}l@{}}1\\ 2\\ 3\\ 4\\ 5\\ 6\\ 7\\ 8\\ 9\\ 10\\ 11\\ 12\\ 13\\ 14\\ 15\end{tabular} &
  \begin{tabular}[c]{@{}l@{}}ct\_state\_ttl\\ sttl\\ dttl\\ ct\_dst\_sport\_ltm\\ ct\_dst\_src\_ltm\\ ct\_src\_dport\_ltm\\ Sload\\ dmeansz\\ ackdat\\ tcprtt\\ synack\\ swin\\ dwin\\ Dload\\ state\end{tabular} &
  \begin{tabular}[c]{@{}l@{}}sttl\\proto\\ dttl\\swin\\ dwin\\ ct\_state\_ttl\\sbytes\\ state\\ dbytes\\ dmeansz\\ Sload\\ Dload\\ Dintpkt\\ smeansz\\ tcprtt\end{tabular} &
  \begin{tabular}[c]{@{}l@{}}sttl\\ct\_state\_ttl\\dttl\\tcprtt\\ ackdat\\ synack\\ ct\_dst\_sport\_ltm\\ ct\_dst\_src\_ltm\\ Sload\\ ct\_src\_dport\_ltm\\ state\\ ct\_srv\_src\\ ct\_srv\_dst\\ ct\_src\_ ltm\\ Sjit\end{tabular} \\ \hline
\end{tabular}
\end{table}

\begin{figure*}[ht]
\begin{subfigure}{.5\textwidth}
  \centering
  \includegraphics[width=8cm, height=3.5cm]{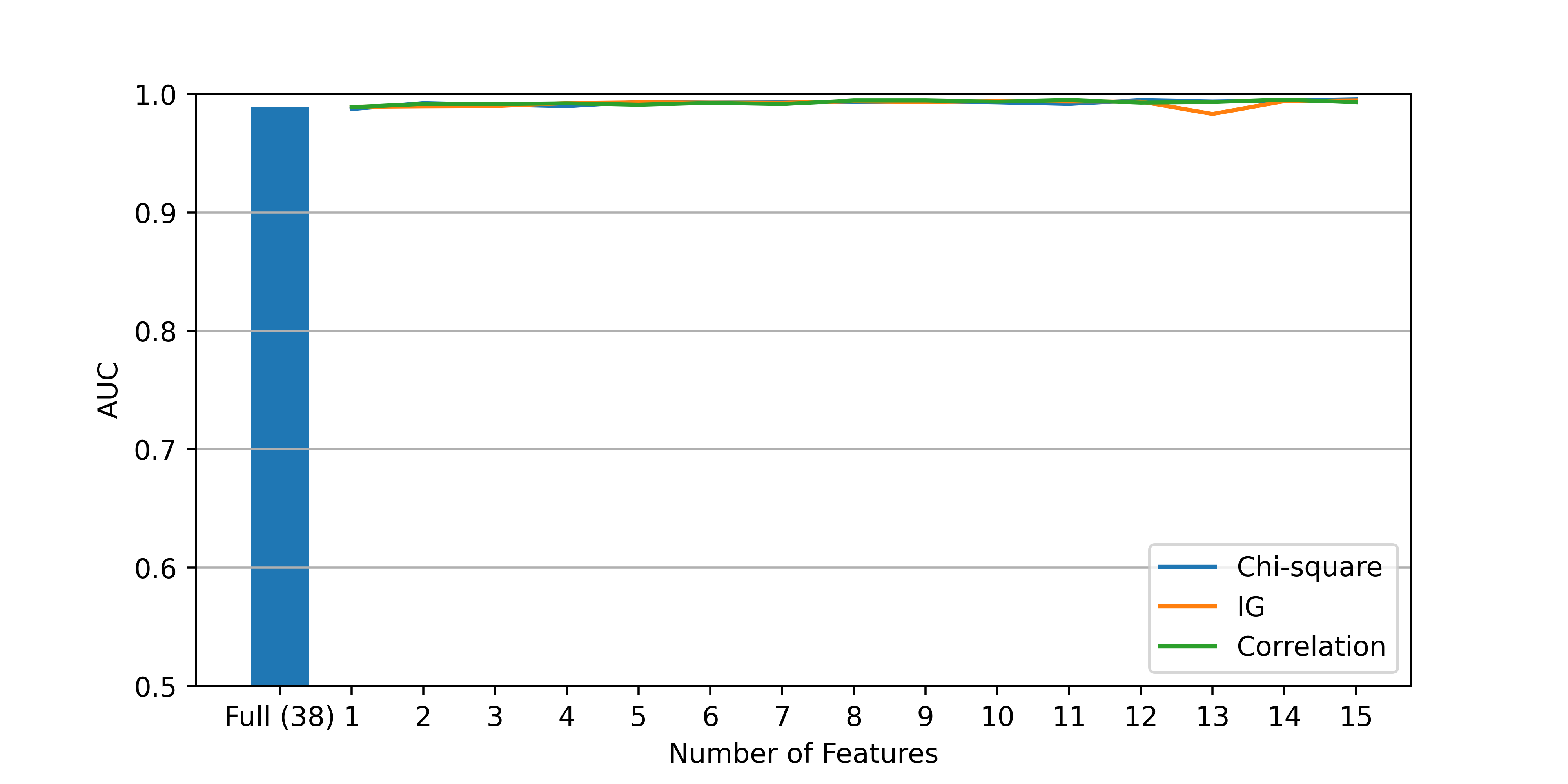}  
  \caption{DFF}
  \label{u}
\end{subfigure}
\hfill
\begin{subfigure}{.5\textwidth}
  \centering
  \includegraphics[width=8cm, height=3.5cm]{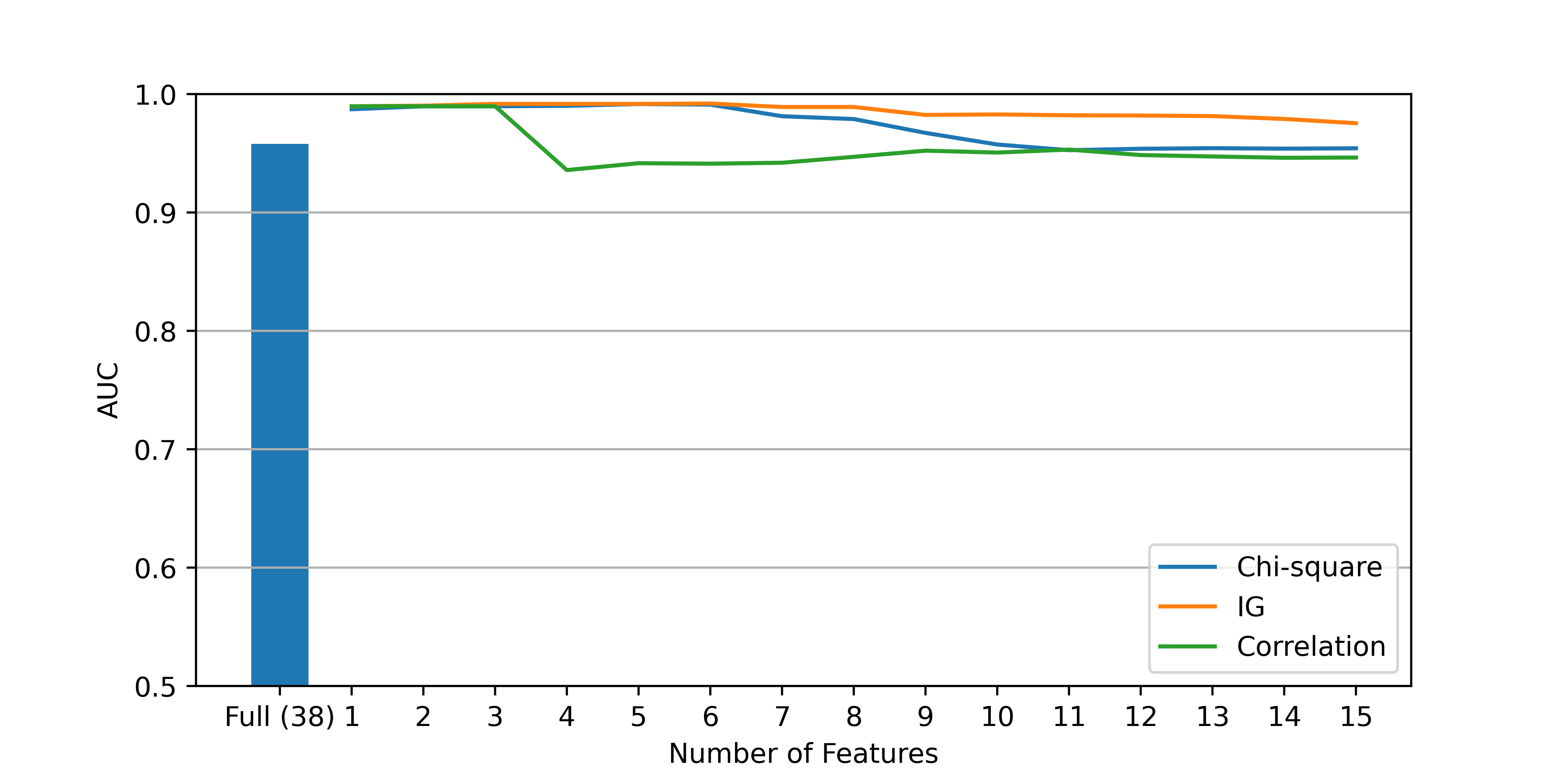}  
  \caption{RF}
  \label{uu}
\end{subfigure}
\caption{UNSW-NB15 detection performance}
\label{tt}
\end{figure*}

\begin{multicols}{2}
In the initial experiment, the features of the UNSW-NB15 dataset have been ranked, and the rankings of the top 15 features for the three respective feature selection algorithms are shown in Table \ref{ft}. Although a direct comparison is difficult due to each dataset's unique choice of features and feature naming, we can observe that there is no clear and consistent ranking. Figure \ref{tt} shows the attack detection performance (y-axis) using the AUC score. The AUC score is used as it is not sensitive to the class imbalance in the NIDS datasets considered. The AUC score is shown for different feature subsets (x-axis) sizes, ranging from 1 to 15. For comparison, the bar on the left represents the AUC results achieved with the full feature set. The three graphs in each figure represent the results based on ranking the features of the three feature selection algorithms. The sub-figure on the left (a) shows the results for the DFF classifier, and the sub-figure on the right (b) shows the corresponding result for the RF classifier.

\begin{figure}[ht]
    \centering
    \includegraphics[width=8cm, height=4cm]{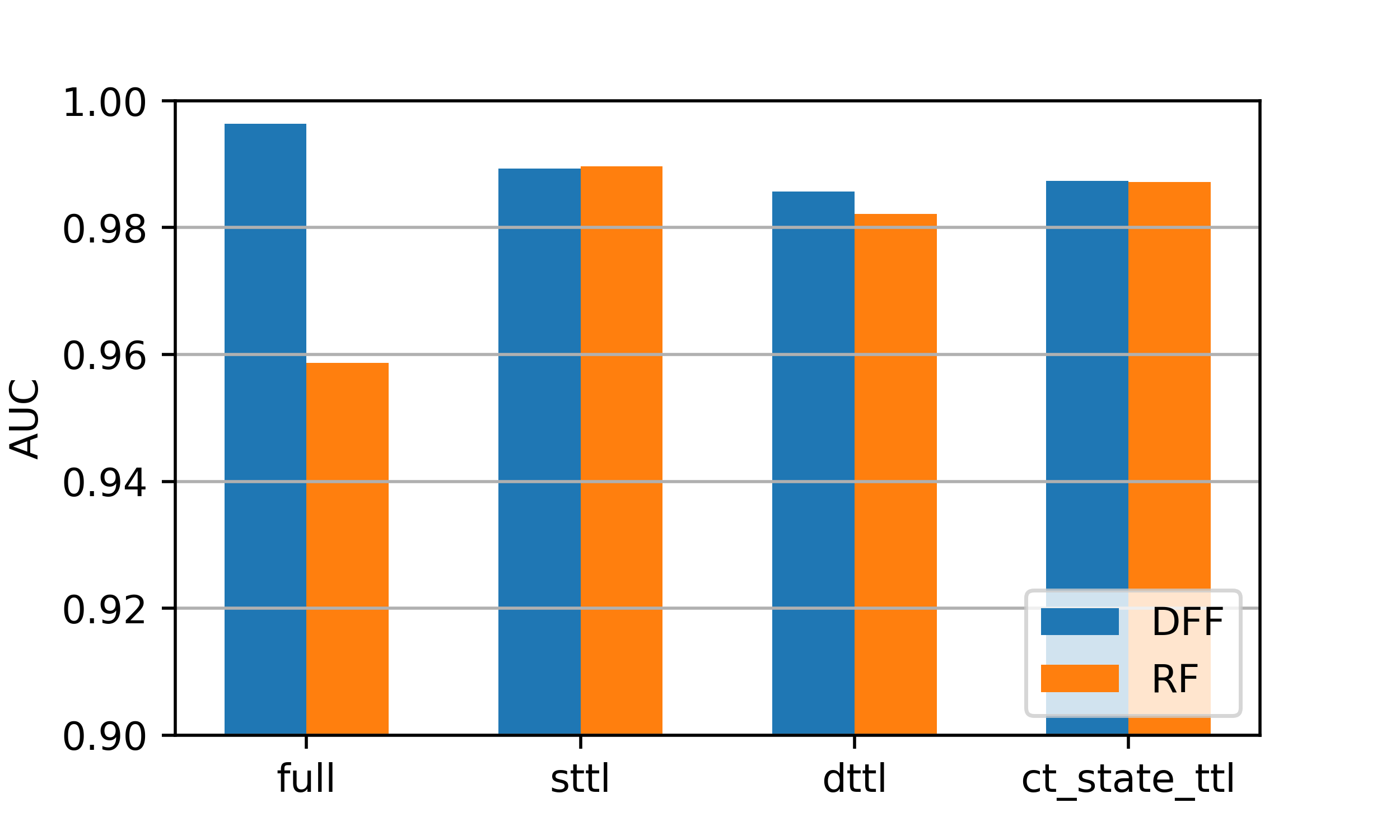}
    \caption{AUC of TTL-based features in UNSW-NB15}
    \label{ttl}
\end{figure}

What immediately stands out in the figure is that the maximum AUC is achieved with only a single feature. Adding any of the next highest-ranked features does not provide any significant additional benefit in classification accuracy. This is consistent for both the DFF and the RF classifier. Further investigation, prompted by this unexpected result, showed that the top-ranked features for the three feature selection algorithms are all Time-to-Live (TTL)-based features in the UNSW-NB15 dataset, ie \textit{sttl}, \textit{dttl} and \textit{ct\_state\_ttl}. A more detailed analysis considered the AUC score achieved by each of these three TTL-based features individually for both classifiers. Figure \ref{ttl} shows these results. As a baseline, the AUC score achieved with the full feature set of 38 features is shown on the left. For the RF classifier, the performance achieved by using any TTL-based features individually is even higher than that of the full features. The addition of irrelevant or `noisy' features that can decrease the classification performance in RF classifiers is not surprising \cite{chandrashekar2014survey}. However, what is indeed surprising is that a single TTL-based feature achieves close to the maximum classification performance achieved by the full feature set of 38 features. We cannot think of any plausible explanation why a single TTL-based feature should have such predictive power for the detection of attack traffic in a practical network scenario. What is a more likely explanation of the observed phenomenon is that it is due to the design of the testbed where the synthetic dataset was generated, e.g., the respective placement of attack and victim nodes.
\end{multicols}

\begin{multicols}{2}

Therefore, we believe that for any reliable and valid evaluation of ML-based classifiers, the three TTL features should be removed from the UNSW-NB15 dataset, as they provide an unrealistically high predictive power, similar to what could be called a 'hidden label'. Although some papers, such as \cite{janarthanan2017feature}, \cite{zhang_wu_gao_wang_xu_liu_2018}, and \cite{al-zewairi_almajali_awajan_2017}, with a focus on feature selection, have reported the \textit{sttl} feature as the most relevant for attack detection, so far no paper has investigated and reported the problem of the improbably high predictive power of a single TTL-based feature, to the best of our knowledge. Since the UNSW-NB15 dataset is highly relevant and widely used in the research community to evaluate ML-based network intrusion detection systems, we believe this is a significant finding. Based on our analysis, using the full UNSW-NB15 feature set, including TTL-based features, as done in many recent research papers, might produce compromised results that are unlikely to generalise to other realistic network scenarios. Consequently, for our further analysis in this paper, we remove the \textit{sttl}, \textit{dttl}, and \textit{ct\_state\_ttl} features from the UNSW-NB15 dataset.

\subsubsection{Revised Feature Set}
\end{multicols}
\begin{table}[ht]\scriptsize
\centering
\caption{UNSW-NB15 (revised feature set) and NF-UNSW-NB15 ranked features}
\label{unswranked}
\begin{tabular}{l|llll|}
\cline{2-5}
 &
  \multicolumn{1}{l|}{\textbf{Rank}} &
  \multicolumn{1}{l|}{\textbf{Chi-square}} &
  \multicolumn{1}{l|}{\textbf{Information Gain}} &
  \textbf{Correlation} \\ \hline
\multicolumn{1}{|l|}{\textbf{\begin{tabular}[c]{@{}l@{}}UNSW-\\ NB15\end{tabular}}} &
  \begin{tabular}[c]{@{}l@{}}1\\ 2\\ 3\\ 4\\ 5\\ 6\\ 7\\ 8\\ 9\\ 10\\ 11\\ 12\\ 13\\ 14\\ 15\end{tabular} &
  \begin{tabular}[c]{@{}l@{}}ct\_dst\_sport\_ltm\\ ct\_dst\_src\_ltm\\ ct\_src\_dport\_ltm\\ Sload\\ dmeansz\\ ackdat\\ tcprtt\\ synack\\ swin\\ dwin\\ Dload\\ state\\ ct\_srv\_src\\ ct\_srv\_dst\\ dtcpb\end{tabular} &
  \begin{tabular}[c]{@{}l@{}}proto\\ swin\\ dwin\\ sbytes\\ state\\ dbytes\\ dmeansz\\ Sload\\ Dload\\ Dintpkt\\ smeansz\\ tcprtt\\ ackdat\\ synack\\ Dpkts\end{tabular} &
  \begin{tabular}[c]{@{}l@{}}tcprtt\\ ackdat\\ synack\\ ct\_dst\_sport\_ltm\\ ct\_dst\_src\_ltm\\ Sload\\ ct\_src\_dport\_ltm\\ state\\ ct\_srv\_src\\ ct\_srv\_dst\\ ct\_src\_ ltm\\ Sjit\\ smeansz\\ ct\_dst\_ltm\\ trans\_depth\end{tabular} \\ \hline
\multicolumn{1}{|l|}{\textbf{\begin{tabular}[c]{@{}l@{}}NF-UNSW-\\ NB15\end{tabular}}} &
  \begin{tabular}[c]{@{}l@{}}1\\ 2\\ 3\\ 4\\ 5\\ 6\\ 7\\ 8\end{tabular} &
  \begin{tabular}[c]{@{}l@{}}PROTOCOL\\ TCP\_FLAGS\\ FLOW\_DURATION\\ IN\_BYTES\\ OUT\_PKTS\\ OUT\_BYTES\\ L7\_PROTO\\ IN\_PKTS\end{tabular} &
  \begin{tabular}[c]{@{}l@{}}TCP\_FLAGS\\ IN\_BYTES\\ OUT\_BYTES\\ OUT\_PKTS\\ IN\_PKTS\\ L7\_PROTO\\ FLOW\_DURATION\\ PROTOCOL\end{tabular} &
  \begin{tabular}[c]{@{}l@{}}PROTOCOL\\ IN\_BYTES\\ L7\_PROTO\\ FLOW\_DURATION\\ IN\_PKTS\\ OUT\_BYTES\\ OUT\_PKTS\\ TCP\_FLAGS\end{tabular} \\ \hline
\end{tabular}
\end{table}
\begin{multicols}{2}
 
\begin{figure*}[ht]
\begin{subfigure}{.5\textwidth}
  \centering
  \includegraphics[width=8cm, height=3.5cm]{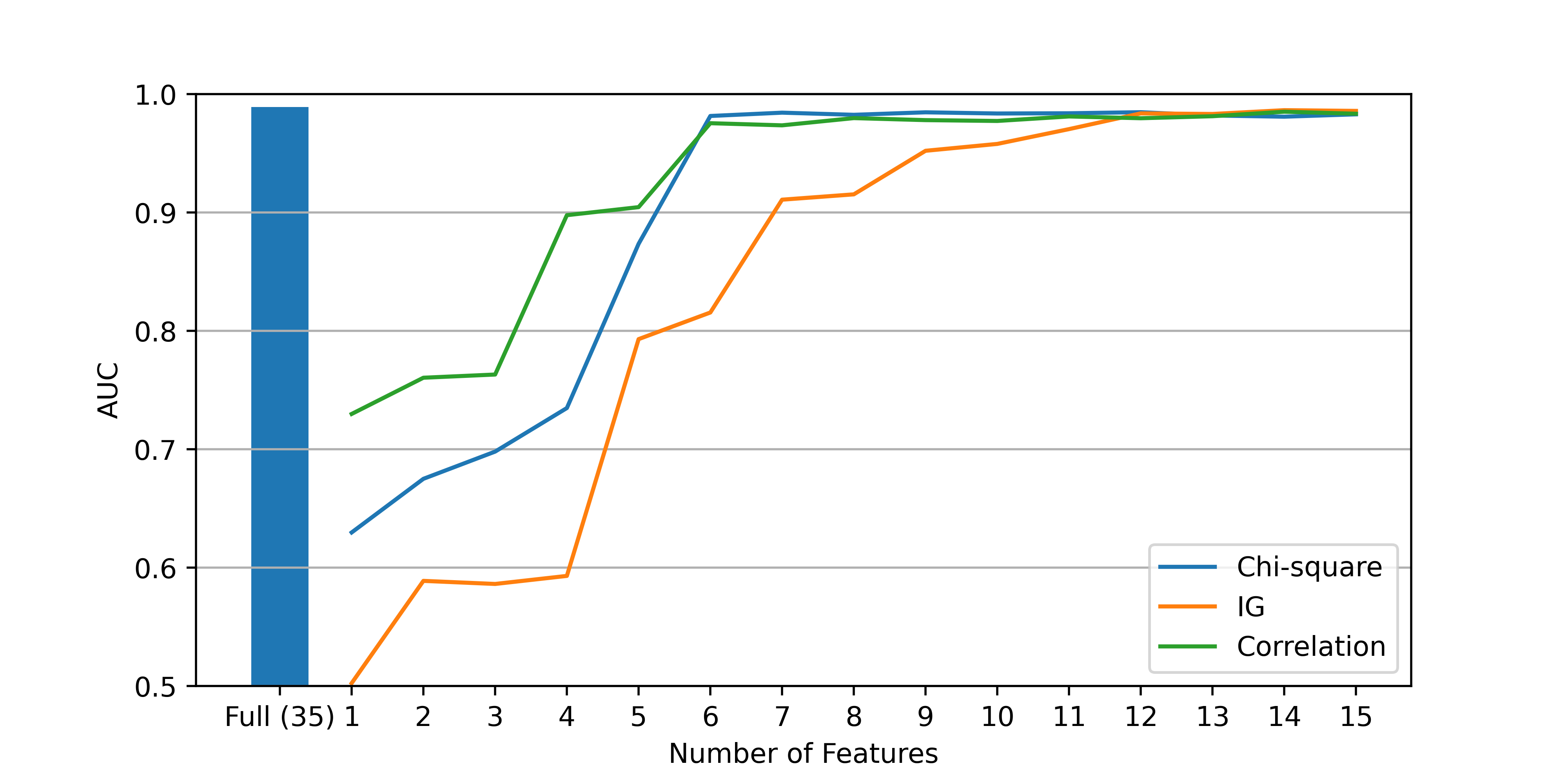}  
  \caption{UNSW-NB15 using DFF}
  \label{fig:dnnunsw}
\end{subfigure}
\hfill
\begin{subfigure}{.5\textwidth}
  \centering
  \includegraphics[width=8cm, height=3.5cm]{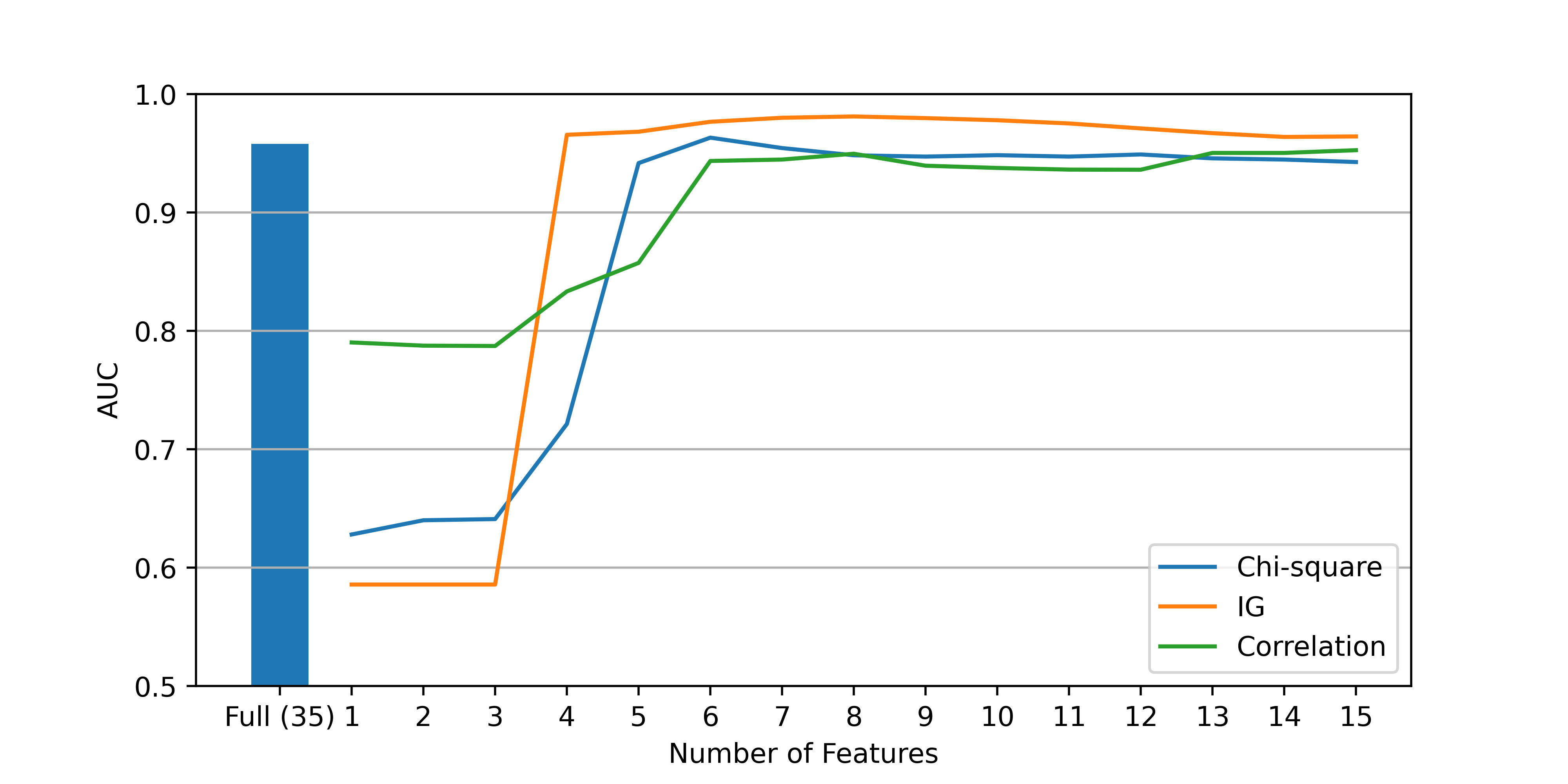}  
  \caption{UNSW-NB15 using RF}
  \label{fig:rfunsw}
\end{subfigure}
\hfill
\begin{subfigure}{.5\textwidth}
  \centering
  \includegraphics[width=8cm, height=3.5cm]{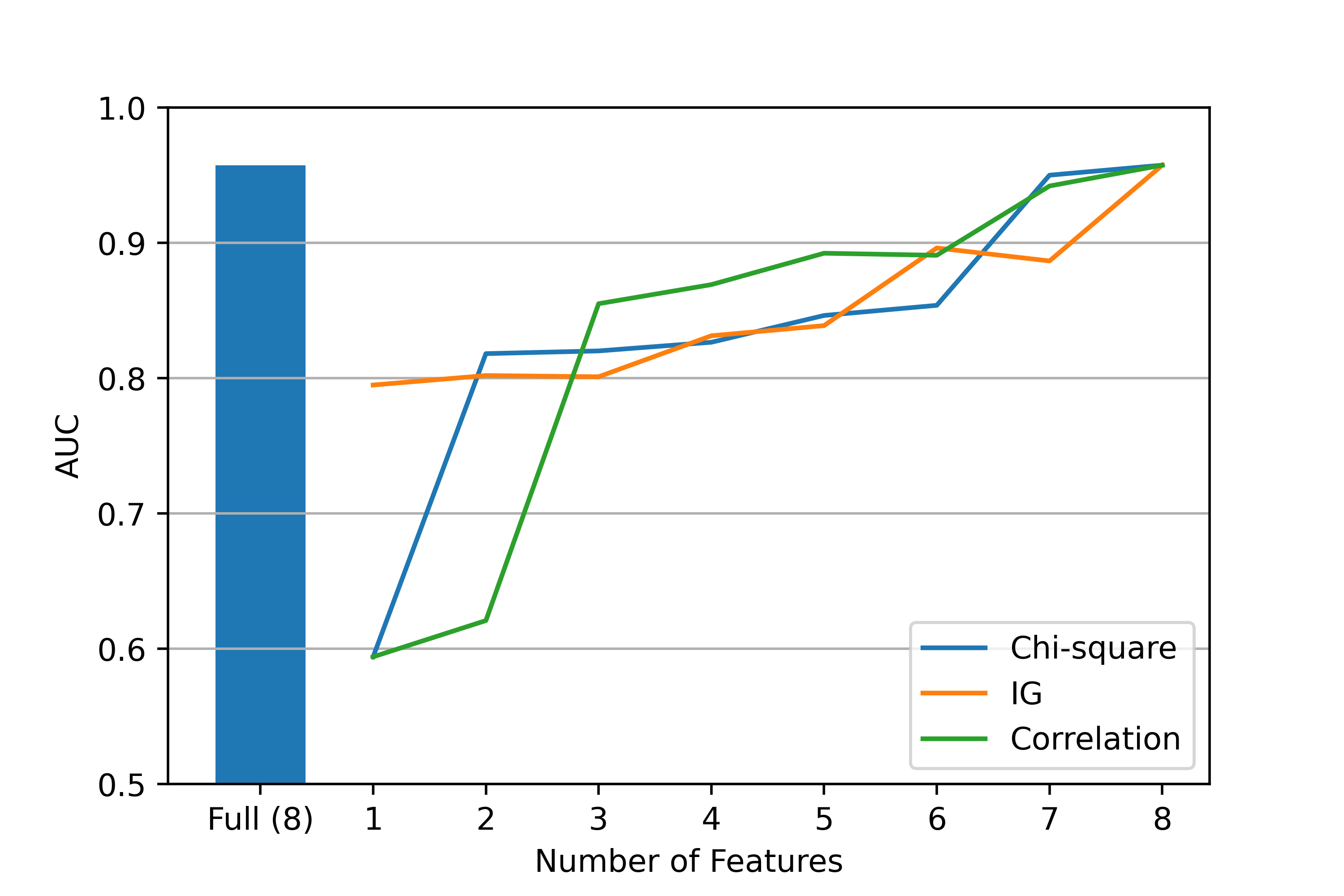}  
  \caption{NF-UNSW-NB15 using DFF}
  \label{fig:dnnunswnf}
\end{subfigure}
\hfill
\begin{subfigure}{.5\textwidth}
  \centering
  \includegraphics[width=8cm, height=3.5cm]{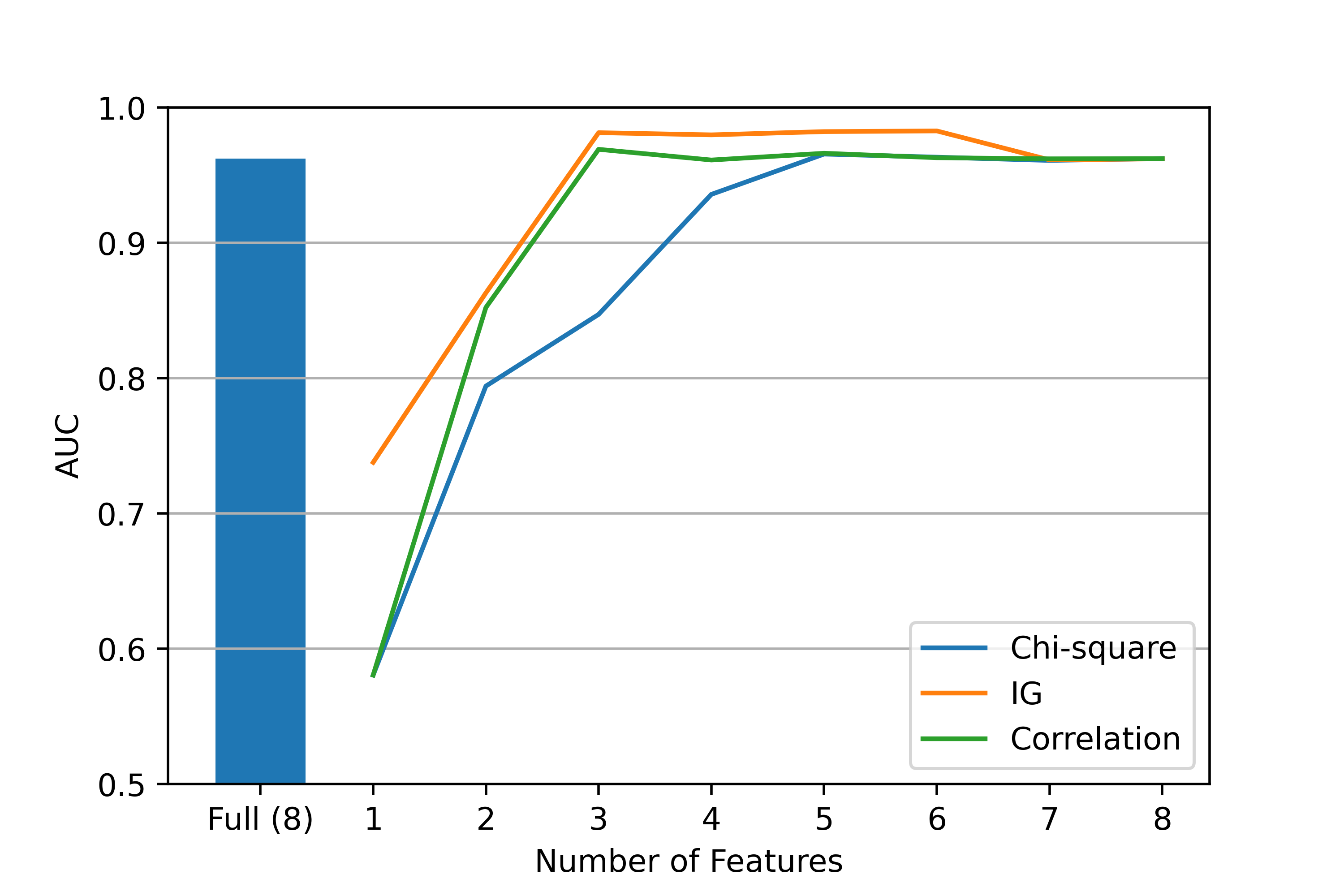}  
  \caption{NF-UNSW-NB15 using RF}
  \label{fig:rfunswnf}
\end{subfigure}
\caption{UNSW-NB15 and NF-UNSW-NB15 detection performance}
\label{fig:fig}
\end{figure*}

Table \ref{unswranked} shows the ranked features for the UNSW-NB15 dataset,  where the three TTL-based features have been removed. The table also shows the feature ranking of the NF-UNSW-NB15 dataset, with a feature set consisting of the following eight NetFlow-based fields: IN\_BYTES, OUT\_BYTES, IN\_PKTS, OUT\_PKTS, PROTOCOL, TCP\_FLAGS, FLOW\_DURATION, and L7\_PROTOCOL. Interestingly, there does not seem to be a consistent feature ranking among the three feature selection algorithms. For example, the top-ranked feature according to IG (\textit{proto}) does not appear in the list of the top 15 features of the other two feature rankings. Figure \ref{fig:fig} shows the attack detection performance (AUC score) on the revised datasets for a feature set size (x-axis) that is increased by increments of 1, with features added in order of their rank according to the feature selection algorithm, i.e., chi-square, IG, and correlation respectively. The top two sub-figures show the results for the UNSW-NB15 dataset and the DFF (a) and RF (b) classifiers, respectively. 

In Figure \ref{fig:fig}, the bar on the left shows the AUC score achieved with the complete set of features, that is, 35 features in this case. This illustrates the importance of choosing a suitable feature selection algorithm. For the DFF classifier, we observe a gradual increase in the AUC score with the addition of features up to 6 features, where the AUC score converges to its maximum value for the chi-square and correlation-based feature ranks. For features ranked according to IG, this convergence to the maximum occurs only for 12 features or more. In the case of the RF classifier (Figure \ref{tt}(b)), we observe a similar pattern, where the maximum AUC score is achieved for a small subset of features of size 4-6. In this case, the IG algorithm performs better and achieves the (close to) maximum AUC score for only four features. Figures \ref{tt} (c) and (d) show the corresponding results for the NF-UNSW-NB15 dataset, where we can observe a similar pattern. However, the difference here is that the range of the x-axis is limited to 1-8 since the dataset contains only eight features.

\end{multicols}
\begin{table}[ht] \scriptsize
\centering
\caption{UNSW-NB15 and NF-UNSW-NB15 full metrics}
\label{unswmetrics}
\begin{tabular}{|c|c|c|r|r|r|r|r|r|}
\hline
\textbf{Dataset} &
  \textbf{Classifier} &
  \textbf{\begin{tabular}[c]{@{}c@{}}Features\\ (Count)\end{tabular}} &
  \multicolumn{1}{c|}{\textbf{Accuracy}} &
  \multicolumn{1}{c|}{\textbf{AUC}} &
  \multicolumn{1}{c|}{\textbf{\begin{tabular}[c]{@{}c@{}}F1\\ Score\end{tabular}}} &
  \multicolumn{1}{c|}{\textbf{DR}} &
  \multicolumn{1}{c|}{\textbf{FAR}} &
  \multicolumn{1}{c|}{\textbf{\begin{tabular}[c]{@{}c@{}}Prediction\\ Time (\textmu s)\end{tabular}}} \\ \hline
\multirow{4}{*}{\textbf{UNSW-NB15}}    & \multirow{2}{*}{\textbf{DFF}} & \textbf{Full (35)} & 96.93\% & 0.9890 & 0.76 & 99.64\% & 3.21\%  & 4.34 \\ \cline{3-9} 
                                      &                               & \textbf{CHI (7)}   & 96.63\% & 0.9843 & 0.74 & 99.14\% & 3.50\%  & 3.84 \\ \cline{2-9} 
                                      & \multirow{2}{*}{\textbf{RF}}  & \textbf{Full (35)} & 99.27\% & 0.9580 & 0.92 & 91.95\% & 0.36\%  & 5.18 \\ \cline{3-9} 
                                      &                               & \textbf{IG (7)}    & 98.62\% & 0.9800 & 0.87 & 97.31\% & 1.31\%  & 4.52 \\ \hline
\multirow{4}{*}{\textbf{NF-UNSW-NB15}} & \multirow{2}{*}{\textbf{DFF}} & \textbf{Full (8)}  & 90.92\% & 0.9574 & 0.57 & 82.38\% & 8.68\% & 4.09 \\ \cline{3-9} 
                                      &                               & \textbf{CHI (7)}  & 94.99\% & 0.9500 & 0.65 & 79.44\% & 4.28\% & 3.63 \\ \cline{2-9} 
                                      & \multirow{2}{*}{\textbf{RF}}  & \textbf{Full (8)}  & 98.50\% & 0.9622 & 0.85 & 93.71\% & 1.27\%  & 4.54 \\ \cline{3-9} 
                                      &                               & \textbf{IG (3)}    & 98.00\% & 0.9814 & 0.81 & 98.28\% & 2.01\%  & 3.99 \\ \hline
\end{tabular}
\end{table}
\begin{multicols}{2}

Table \ref{unswmetrics} shows the key classification evaluation metrics for both the complete set of features and the size of the subset of features (with the corresponding feature selection algorithm) that achieves the highest classification performance in terms of the AUC score. The results are shown for the original UNSW-NB15 and NF-UNSW-NB15 datasets. What is consistent across all different cases and metrics is that the chosen feature subset, with a fraction of the full size (in the case of UNSW-NB15), achieves either better or close-to-equal classification performance compared to the point where the complete feature set is used. This means that by using a significantly reduced feature set, excellent classification performance can be achieved but with reduced resource demand for feature extraction and storage considerably, as well as significantly reduced model complexity and prediction times. This is particularly relevant for scenarios with resource-constrained edge and IoT devices, particularly in the case of IoT. 



\subsection{ToN-IoT and NF-ToN-IoT}
\end{multicols}
\begin{table}[ht]\scriptsize
\centering
\caption{ToN-IoT and NF-ToN-IoT ranked features}
\label{tonranked}
\begin{tabular}{l|llll|}
\cline{2-5}
 &
  \multicolumn{1}{l|}{\textbf{Rank}} &
  \multicolumn{1}{l|}{\textbf{Chi-square}} &
  \multicolumn{1}{l|}{\textbf{Information Gain}} &
  \textbf{Correlation} \\ \hline
\multicolumn{1}{|l|}{\textbf{\begin{tabular}[c]{@{}l@{}}ToN-\\ IoT\end{tabular}}} &
  \begin{tabular}[c]{@{}l@{}}1\\ 2\\ 3\\ 4\\ 5\\ 6\\ 7\\ 8\\ 9\\ 10\\ 11\\ 12\\ 13\\ 14\\ 15\end{tabular} &
  \begin{tabular}[c]{@{}l@{}}dns\_query\\ dns\_qclass\\ dns\_rejected\\ proto\\ dns\_RA\\ dns\_RD\\ dns\_qtype\\ dns\_AA\\ dns\_rcode\\ service\\ state\\ http\_status\_code\\ dst\_ip\_bytes\\ src\_bytes\\ dst\_bytes\end{tabular} &
  \begin{tabular}[c]{@{}l@{}}http\_resp\_mime\_types\\ ssl\_subject\\ ssl\_issuer\\ state\\ http\_method\\ weird\_addl\\ ssl\_version\\ weird\_name\\ ssl\_cipher\\ http\_user\_agent\\ service\\ src\_ip\_bytes\\ dst\_ip\_bytes\\ src\_pkts\\ src\_bytes\end{tabular} &
  \begin{tabular}[c]{@{}l@{}}dns\_rejected\\ dns\_RA\\ dns\_RD\\ dns\_AA\\ weird\_notice\\ state\\ service\\ ssl\_established\\ duration\\ ssl\_issuer\\ ssl\_subject\\ src\_bytes\\ http\_method\\ http\_status\_code\\ ssl\_cipher\end{tabular} \\ \hline
\multicolumn{1}{|l|}{\textbf{\begin{tabular}[c]{@{}l@{}}NF-ToN-\\ IoT\end{tabular}}} &
  \begin{tabular}[c]{@{}l@{}}1\\ 2\\ 3\\ 4\\ 5\\ 6\\ 7\\ 8\end{tabular} &
  \begin{tabular}[c]{@{}l@{}}L7\_PROTO\\ FLOW\_DURATION\\ TCP\_FLAGS\\ PROTOCOL\\ IN\_BYTES\\ IN\_PKTS\\ OUT\_PKTS\\ OUT\_BYTES\end{tabular} &
  \begin{tabular}[c]{@{}l@{}}OUT\_BYTES\\ IN\_BYTES\\ TCP\_FLAGS\\ IN\_PKTS\\ L7\_PROTO\\ OUT\_PKTS\\ FLOW\_DURATION\\ PROTOCOL\end{tabular} &
  \begin{tabular}[c]{@{}l@{}}TCP\_FLAGS\\ OUT\_BYTES\\ OUT\_PKTS\\ IN\_PKTS\\ IN\_BYTES\\ FLOW\_DURATION\\ PROTOCOL\\ L7\_PROTO\end{tabular} \\ \hline
\end{tabular}
\end{table}
\begin{multicols}{2}

\begin{figure*}[ht]
\begin{subfigure}{.5\textwidth}
  \centering
  \includegraphics[width=8cm, height=3.5cm]{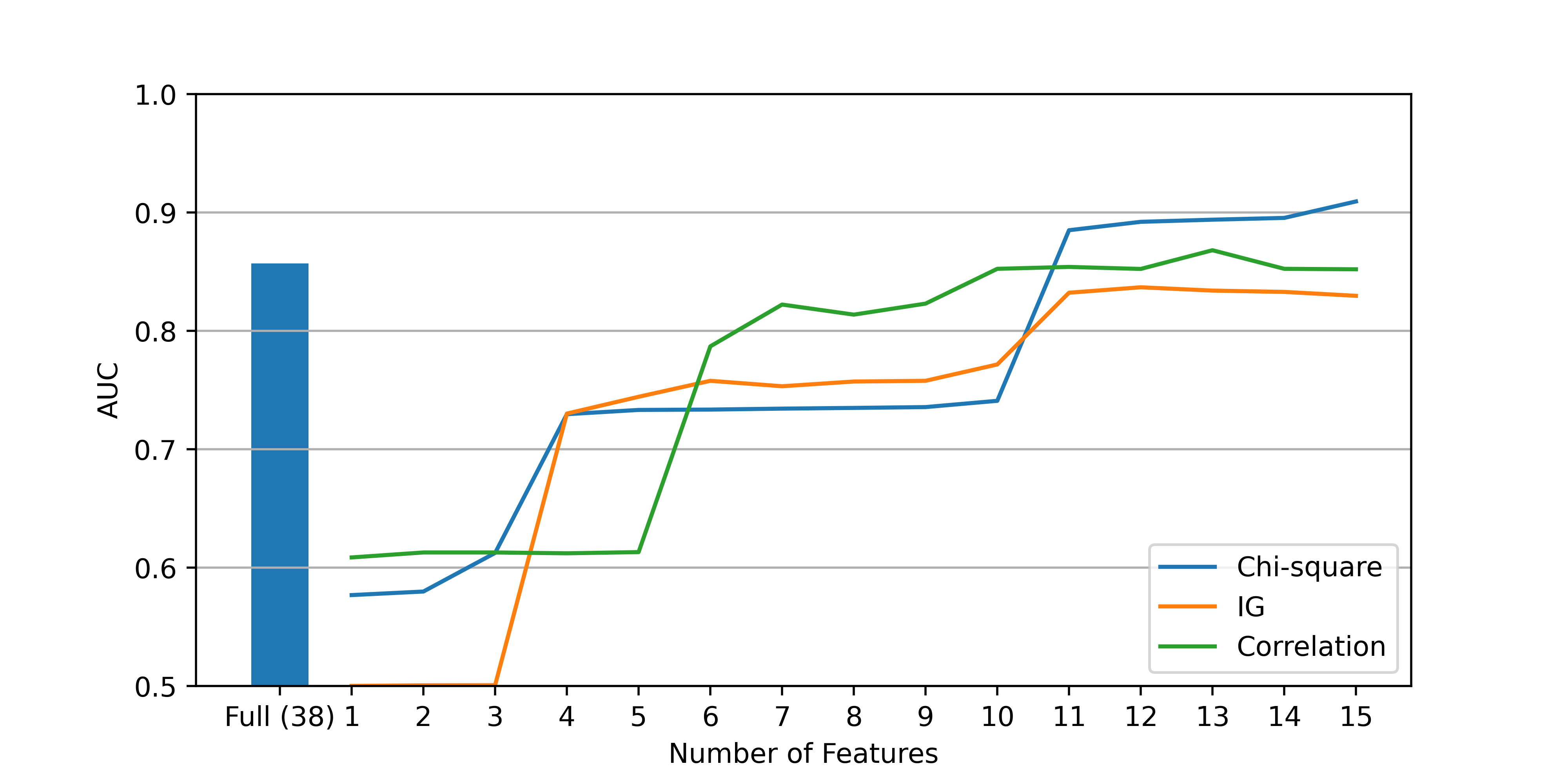}  
  \caption{ToN-IoT using DFF}
  \label{fig:dnnton}
\end{subfigure}
\hfill
\begin{subfigure}{.5\textwidth}
  \centering
  \includegraphics[width=8cm, height=3.5cm]{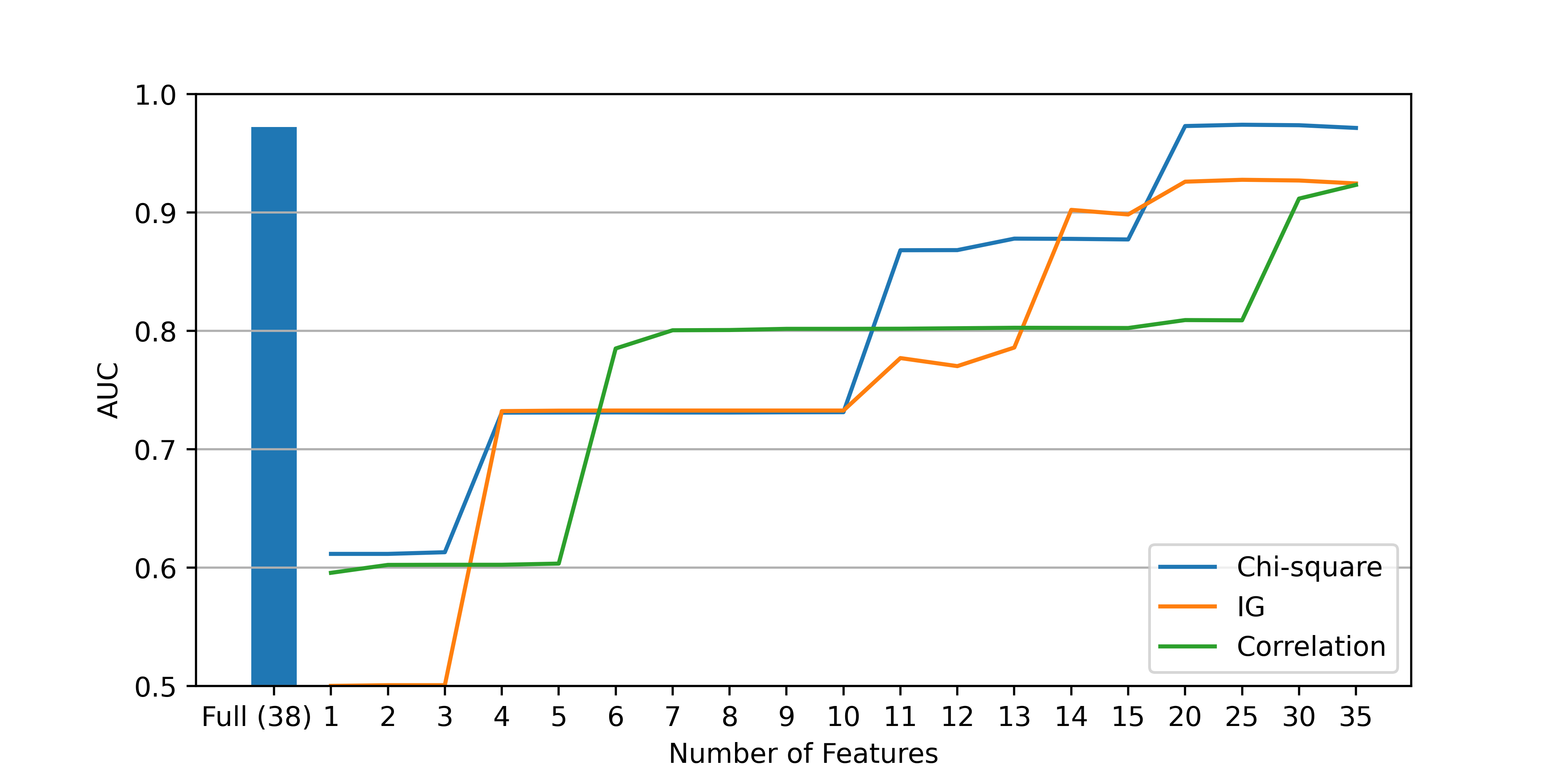}  
  \caption{ToN-IoT using RF}
  \label{fig:rfton}
\end{subfigure}
\hfill
\begin{subfigure}{.5\textwidth}
  \centering
  \includegraphics[width=8cm, height=3.5cm]{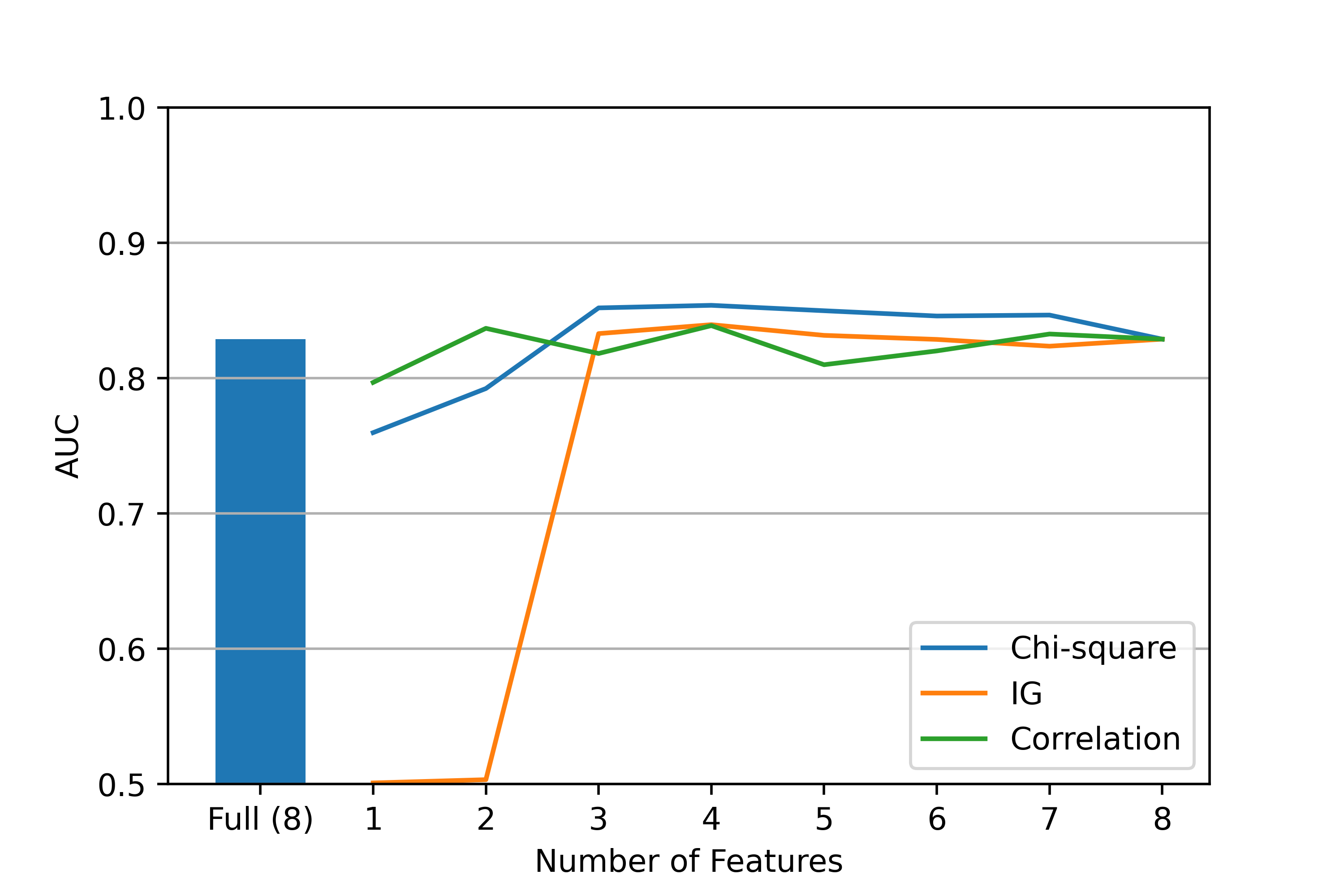}  
  \caption{NF-ToN-IoT using DFF}
  \label{fig:dnntonnf}
\end{subfigure}
\hfill
\begin{subfigure}{.5\textwidth}
  \centering
  \includegraphics[width=8cm, height=3.5cm]{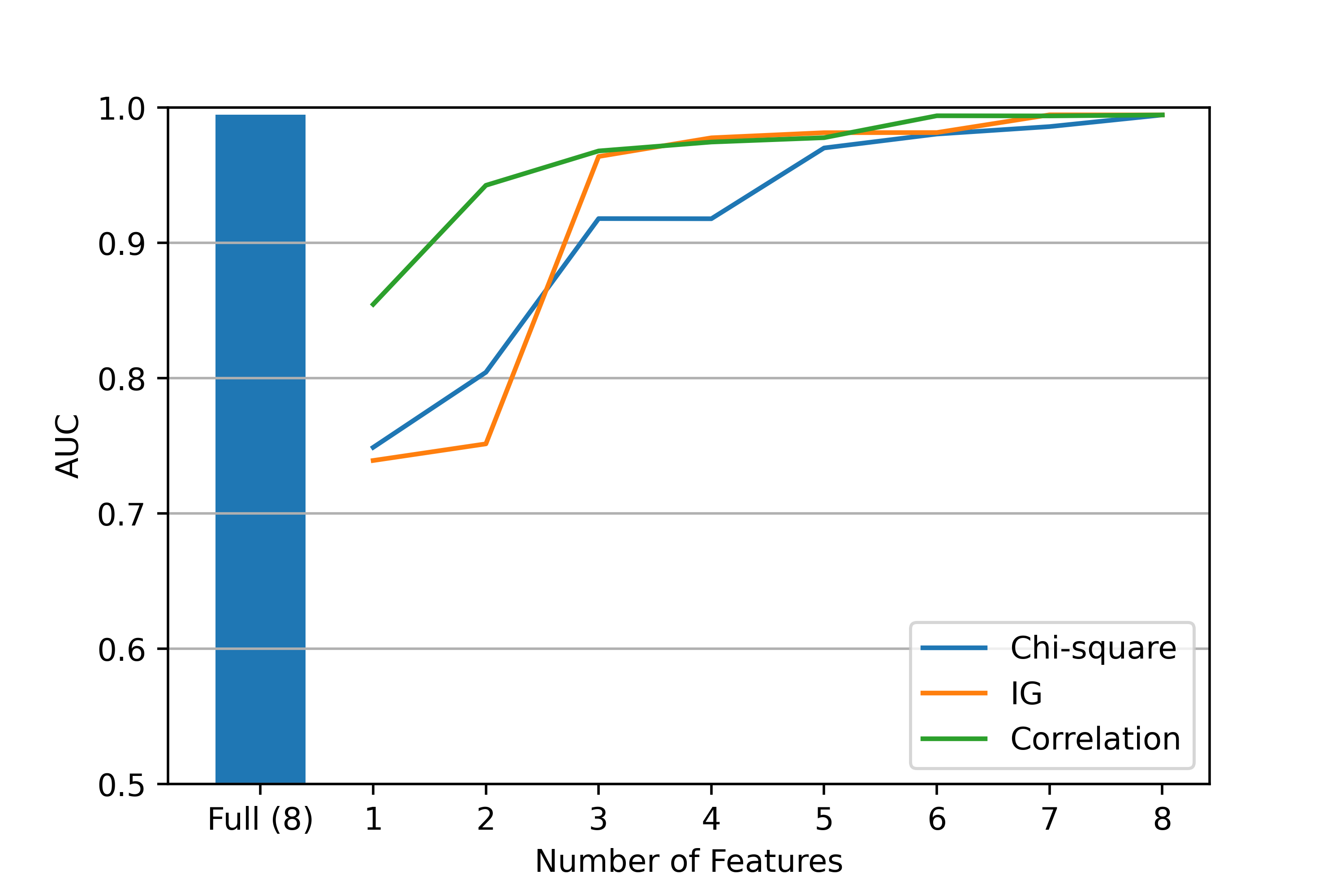}  
  \caption{NF-ToN-IoT using RF}
  \label{fig:rftonnf}
\end{subfigure}
\caption{ToN-IoT and NF-ToN-IoT detection performance}
\label{fig:fig2}
\end{figure*}

Table \ref{tonranked} shows the ranking features for the ToN-IoT dataset and its NetFlow-based variant NF-ToN-IoT. Figure \ref{fig:fig2} shows the attack detection performance (AUC score) for these datasets using subsets of ranked features or increasing size. As in the case of the UNSW-NB15 dataset, we see significant differences in feature ranks among the three feature selection algorithms. For both classifiers and the ToN-IoT dataset, we see a more gradual conversion to the maximum AUC score, and a higher number of features ($>$10) is required to converge to the maximum compared to the UNSW-NB15 dataset, where 4-6 features were sufficient. In contrast, for the NF-ToN-IoT dataset, a relatively small fraction of 2-3 features (of the total of 8) can achieve a close to maximum performance. Again, we notice a significant difference between the three feature selection algorithms.


Table \ref{tonmetrics} shows the full evaluation metrics of the ToN-IoT and NF-ToN-IoT datasets. As already observed in Figure \ref{fig:fig2}, In the case of ToN-IoT, a much larger feature sub-set is required to achieve the maximum classification performance, i.e., 15 features in the case of DFF and 20 features in the case of RF. However, the feature sub-sets achieve slightly better performance and lower prediction time than the complete set with 38 features in both cases. A similar result is observed for the RF classifier and the NF-ToN-IoT dataset, where 6 out of 8 features are required to achieve the maximum AUC score. In contrast, the DNN classifier only requires the top three features based on the chi-square ranking to achieve the best performance, which happens to be better than the entire set of features. In summary, there seems to be no one-size-fits-all solution, and the behaviour is quite different for different datasets, feature selection algorithms, and classifiers. 
\end{multicols}
\begin{table}[ht] \scriptsize
\centering
\caption{ToN-IoT and NF-ToN-IoT full metrics}
\label{tonmetrics}
\begin{tabular}{|c|c|c|r|r|r|r|r|r|}
\hline
\textbf{Dataset} &
  \textbf{Classifier} &
  \textbf{\begin{tabular}[c]{@{}c@{}}Features\\ (Count)\end{tabular}} &
  \multicolumn{1}{c|}{\textbf{Accuracy}} &
  \multicolumn{1}{c|}{\textbf{AUC}} &
  \multicolumn{1}{c|}{\textbf{\begin{tabular}[c]{@{}c@{}}F1\\ Score\end{tabular}}} &
  \multicolumn{1}{c|}{\textbf{DR}} &
  \multicolumn{1}{c|}{\textbf{FAR}} &
  \multicolumn{1}{c|}{\textbf{\begin{tabular}[c]{@{}c@{}}Prediction\\ Time (\textmu s)\end{tabular}}} \\ \hline
\multirow{4}{*}{\textbf{ToN-IoT}}    & \multirow{2}{*}{\textbf{DFF}} & \textbf{Full (38)} & 96.10\% & 0.8571 & 0.98 & 97.18\% & 34.76\% & 4.86 \\ \cline{3-9} 
                                     &                               & \textbf{CHI (15)}  & 96.58\% & 0.9093 & 0.98 & 97.49\% & 29.31\% & 3.06  \\ \cline{2-9} 
                                     & \multirow{2}{*}{\textbf{RF}}  & \textbf{Full (38)} & 97.35\% & 0.9721 & 0.99 & 97.36\% & 2.94\%  & 5.94  \\ \cline{3-9} 
                                     &                               & \textbf{CHI (20)}   & 97.49\% & 0.9730 & 0.99 & 97.51\% & 2.90\% & 4.32  \\ \hline
\multirow{4}{*}{\textbf{NF-ToN-IoT}} & \multirow{2}{*}{\textbf{DFF}} & \textbf{Full (8)}  & 75.58\% & 0.8306 & 0.83 & 74.13\% & 18.50\% & 3.98  \\ \cline{3-9} 
                                     &                               & \textbf{CHI (3)}   & 85.61\% & 0.8519 & 0.91 & 88.09\% & 24.58\% & 3.03  \\ \cline{2-9} 
                                     & \multirow{2}{*}{\textbf{RF}}  & \textbf{Full (8)}  & 99.38\% & 0.9946 & 1.00 & 99.33\% & 0.42\%  & 5.56  \\ \cline{3-9} 
                                     &                               & \textbf{COR (6)}    & 99.38\% & 0.9946 & 1.00 & 99.33\% & 0.40\%  & 5.06  \\ \hline
\end{tabular}
\end{table}
\begin{multicols}{2}
\subsection{CSE-CIC-IDS2018 and NF-CSE-CIC-IDS2018}
\begin{figure*}[ht]
\begin{subfigure}{.5\textwidth}
  \centering
  \includegraphics[width=8cm, height=3.5cm]{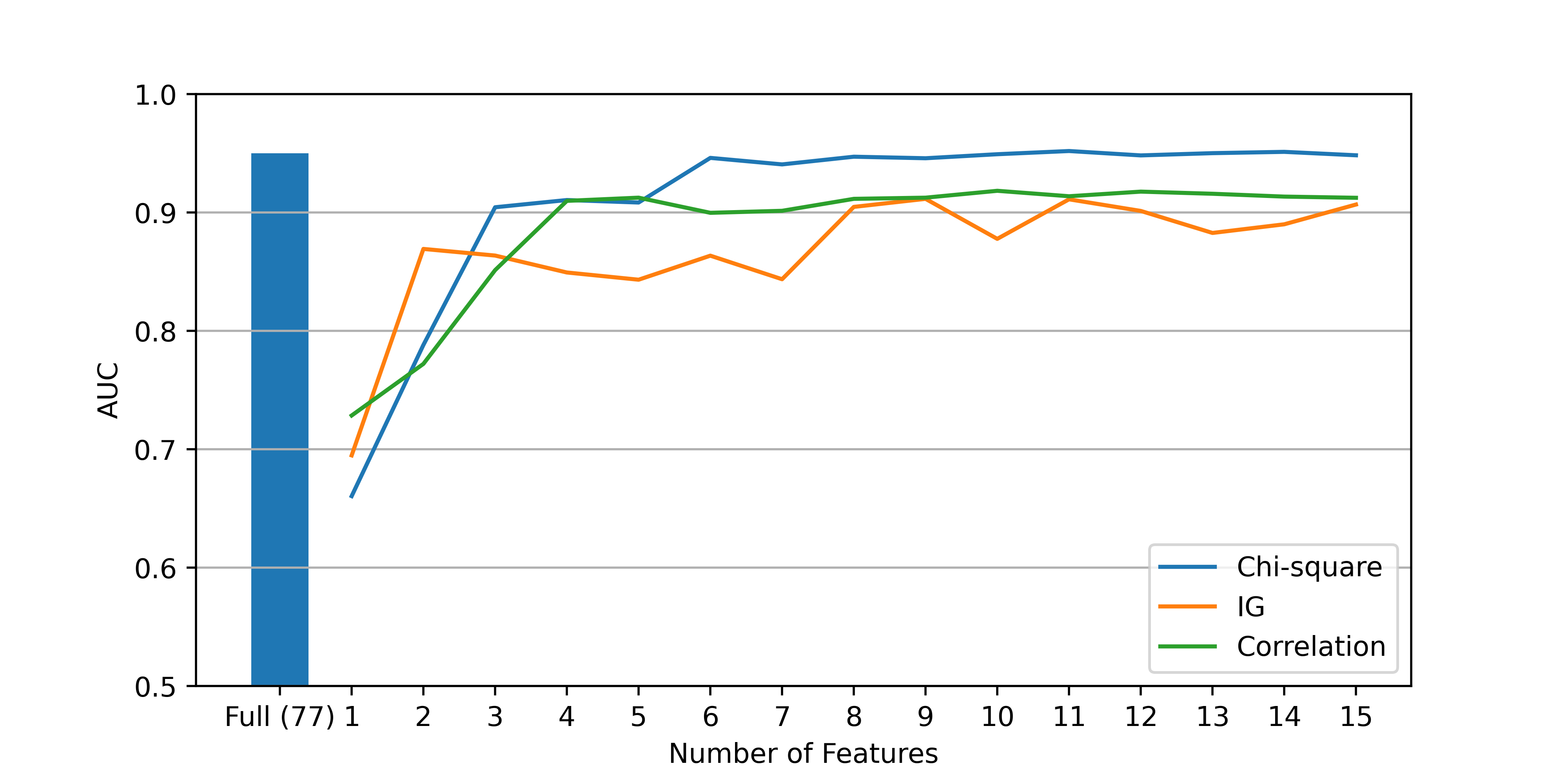}  
  \caption{CSE-CIC-IDS2018 using DFF}
  \label{fig:dnncic}
\end{subfigure}
\hfill
\begin{subfigure}{.5\textwidth}
  \centering
  \includegraphics[width=8cm, height=3.5cm]{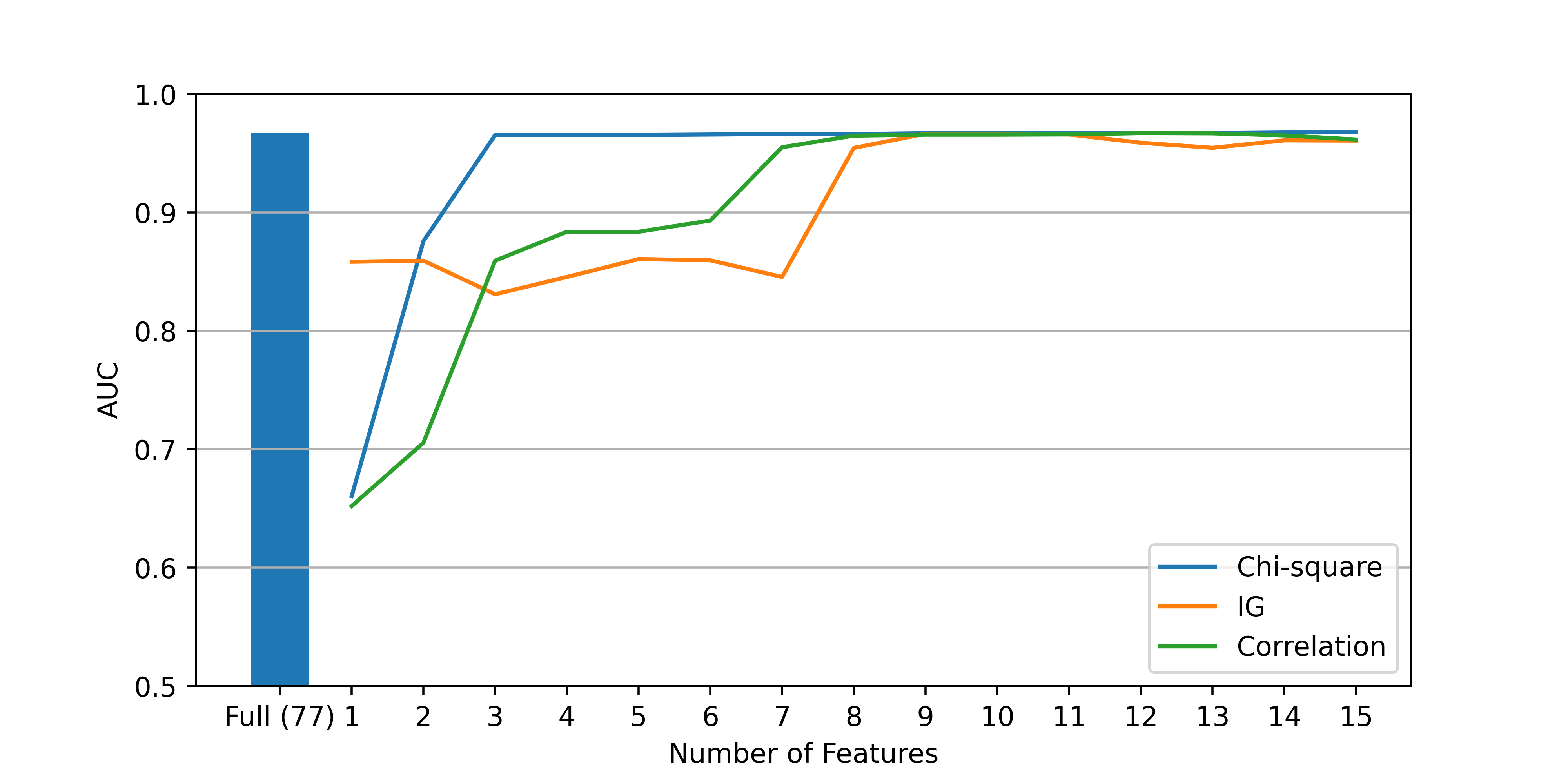}  
  \caption{CSE-CIC-IDS2018 using RF}
  \label{fig:rfcic}
\end{subfigure}
\hfill
\begin{subfigure}{.5\textwidth}
  \centering
  \includegraphics[width=8cm, height=3.5cm]{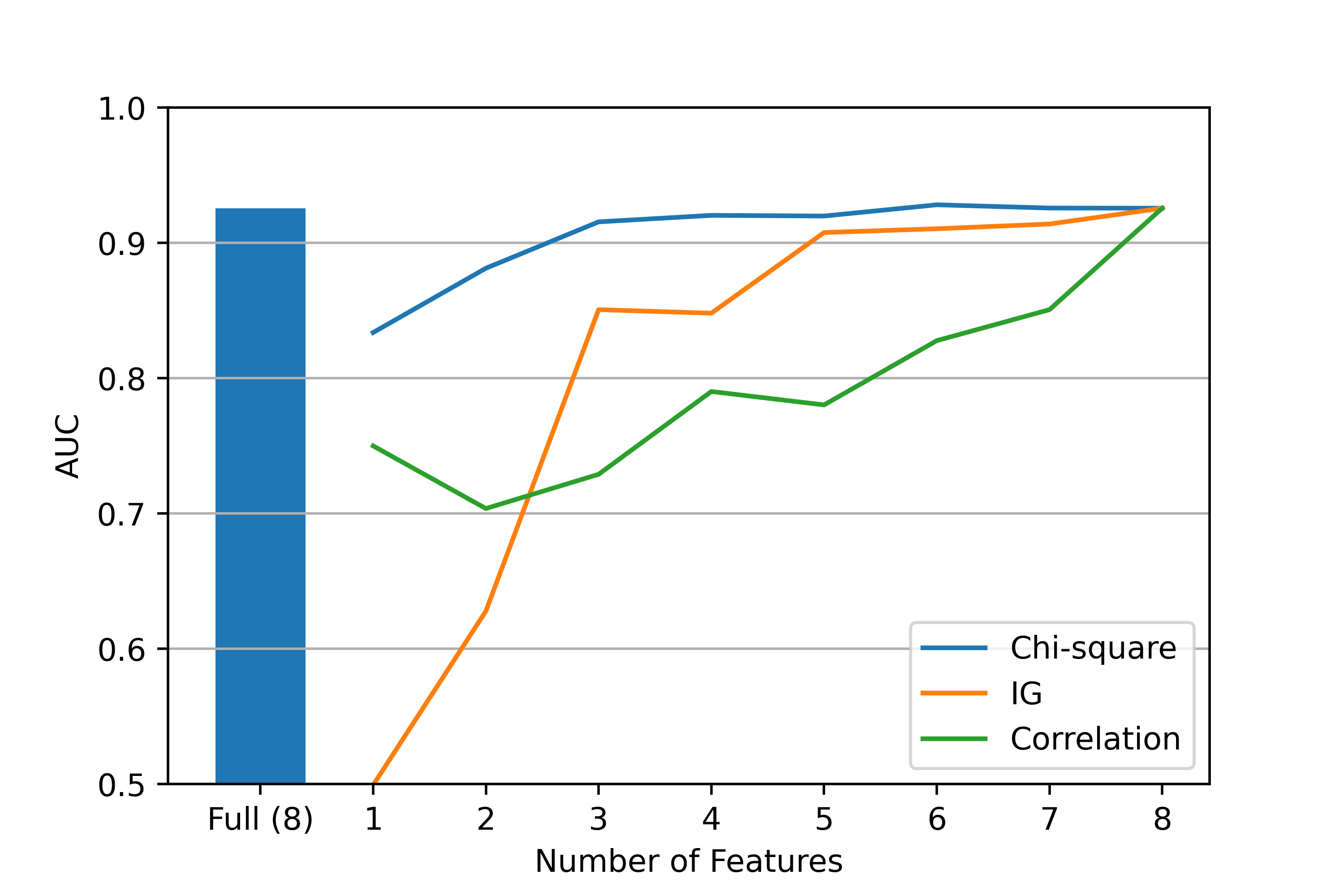}  
  \caption{NF-CSE-CIC-IDS2018 using DFF}
  \label{fig:dnncicnf}
\end{subfigure}
\hfill
\begin{subfigure}{.5\textwidth}
  \centering
  \includegraphics[width=8cm, height=3.5cm]{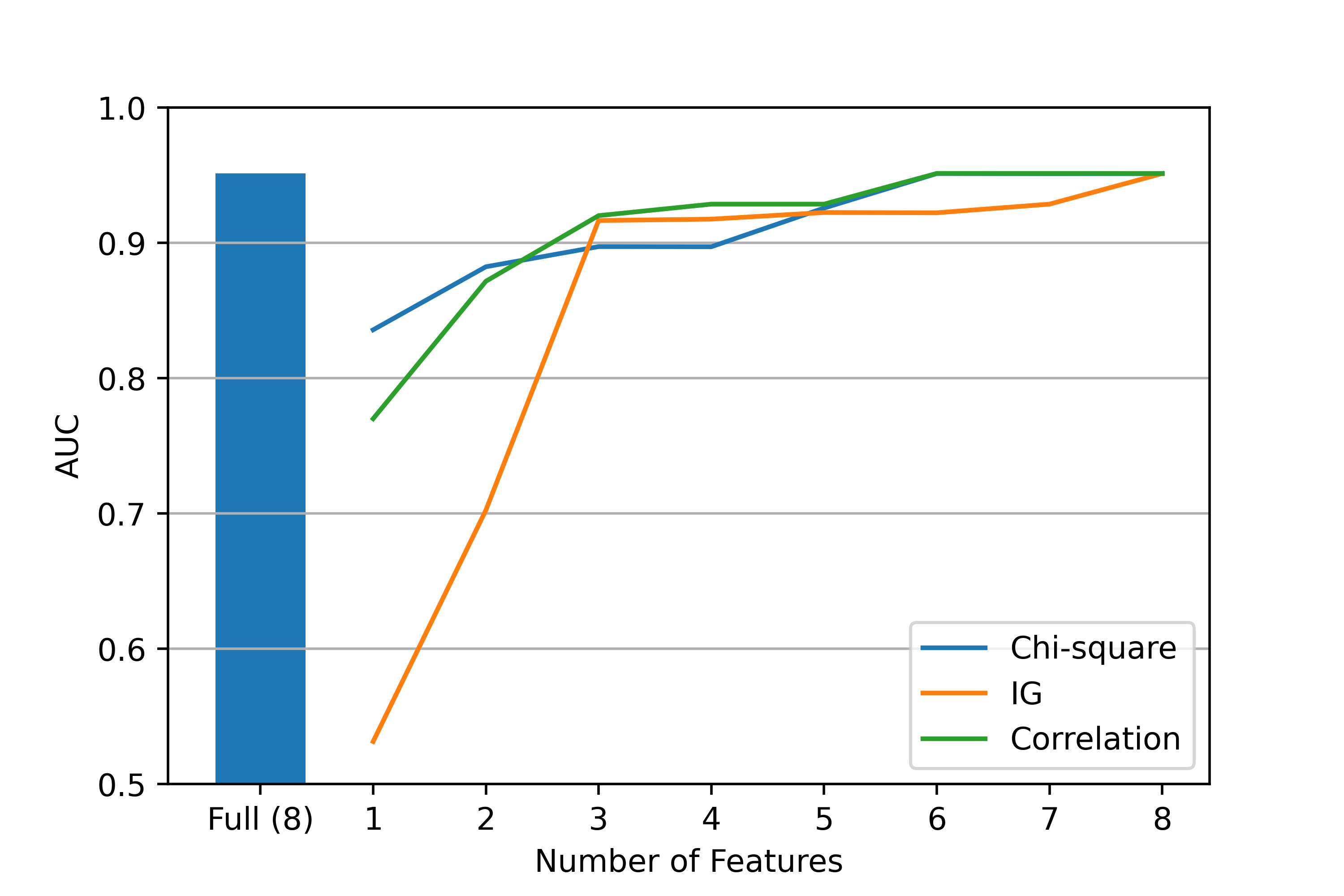}  
  \caption{NF-CSE-CIC-IDS2018 using RF}
  \label{fig:rfcicnf}
\end{subfigure}
\caption{CSE-CIC-IDS2018 and NF-CSE-CIC-IDS2018 detection performance}
\label{fig:fig3}
\end{figure*}

\end{multicols}
\begin{table}[ht]\scriptsize
\centering
\caption{CSE-CIC-IDS2018 and NF-CSE-CIC-IDS2018 ranked features}
\label{cicranked}
\begin{tabular}{l|llll|}
\cline{2-5}
 &
  \multicolumn{1}{l|}{\textbf{Rank}} &
  \multicolumn{1}{l|}{\textbf{Chi-square}} &
  \multicolumn{1}{l|}{\textbf{Information Gain}} &
  \textbf{Correlation} \\ \hline
\multicolumn{1}{|l|}{\textbf{\begin{tabular}[c]{@{}l@{}}CSE-CIC-\\ IDS2018\end{tabular}}} &
  \begin{tabular}[c]{@{}l@{}}1\\ 2\\ 3\\ 4\\ 5\\ 6\\ 7\\ 8\\ 9\\ 10\\ 11\\ 12\\ 13\\ 14\\ 15\end{tabular} &
  \begin{tabular}[c]{@{}l@{}}ACK Flag Cnt\\ Init Bwd Win Byts\\ Init Fwd Win Byts\\ Bwd IAT Tot\\ Protocol\\ Fwd Seg Size Min\\ Fwd PSH Flags\\ SYN Flag Cnt\\ RST Flag Cnt\\ ECE Flag Cnt\\ Bwd Pkt Len Min\\ Bwd IAT Max\\ URG Flag Cnt\\ Fwd Pkt Len Min\\ Pkt Len Min\end{tabular} &
  \begin{tabular}[c]{@{}l@{}}Init Fwd Win Byts\\ Fwd Seg Size Min\\ Fwd IAT Max\\ Fwd IAT Tot\\ Fwd IAT Mean\\ Protocol\\ Flow IAT Max\\ Fwd Seg Size Avg\\ Fwd Pkt Len Mean\\ Flow IAT Mean\\ Tot Len Fwd Pkts\\ Subflow Fwd Byts\\ Flow Duration\\ Fwd Header Len\\ Fwd Pkt Len Max\end{tabular} &
  \begin{tabular}[c]{@{}l@{}}Fwd Seg Size Min\\ ACK Flag Cnt\\ Init Fwd Win Byts\\ RST Flag Cnt\\ ECE Flag Cnt\\ Bwd Pkts/s\\ Fwd Act Data Pkts\\ Tot Fwd Pkts\\ Subflow Fwd Pkts\\ Fwd Header Len\\ Subflow Fwd Byts\\ Tot Len Fwd Pkts\\ Fwd URG Flags\\ CWE Flag Count\\ Fwd IAT Min\end{tabular} \\ \hline
\multicolumn{1}{|l|}{\textbf{\begin{tabular}[c]{@{}l@{}}NF-CSE-CIC-\\ IDS2018\end{tabular}}} &
  \begin{tabular}[c]{@{}l@{}}1\\ 2\\ 3\\ 4\\ 5\\ 6\\ 7\\ 8\end{tabular} &
  \begin{tabular}[c]{@{}l@{}}FLOW\_DURATION\\ L7\_PROTO\\ TCP\_FLAGS\\ PROTOCOL\\ IN\_PKTS\\ IN\_BYTES\\ OUT\_PKTS\\ OUT\_BYTES\end{tabular} &
  \begin{tabular}[c]{@{}l@{}}IN\_BYTES\\ PROTOCOL\\ FLOW\_DURATION\\ OUT\_BYTES\\ TCP\_FLAGS\\ OUT\_PKTS\\ IN\_PKTS\\ L7\_PROTO\end{tabular} &
  \begin{tabular}[c]{@{}l@{}}TCP\_FLAGS\\ IN\_PKTS\\ IN\_BYTES\\ OUT\_BYTES\\ OUT\_PKTS\\ L7\_PROTO\\ PROTOCOL\\ FLOW\_DURATION\end{tabular} \\ \hline
\end{tabular}
\end{table}
\begin{multicols}{2}
Table \ref{cicranked} lists the selected features ranked by their respective scores by each feature selection algorithm for the CSE-CIC-IDS2018 and NF-CSE-CIC-IDS2018 datasets. The top identified features were unique to each applied feature selection algorithm. Figure \ref{fig:fig3} visually represents the attack detection performance of the datasets with the increasing size of the feature set. CSE-CIC-IDS2018 is a large dataset consisting of 77 features. However, a minimal number is sufficient for DFF and RF classifiers to achieve an attack detection performance close to the full feature set. Therefore, the results are similar to the case of the UNSW-NB15 dataset. 

The full evaluation metrics of the CSE-CIC-IDS2018 and NF-CSE-CIC-IDS2018 datasets are listed in Table \ref{cicmetrics}. The table illustrates that most of the original features of the CSE-CIC-IDS2018 dataset can be considered irrelevant, as only about 7\% of them are needed to achieve the maximum attack detection performance resulting in lower inference times. The features selected by the chi-square technique best detect the attacks present in these datasets. The top six and three features are required for the DFF and RF classifiers to achieve maximum performance in the CSE-CIC-IDS2018 dataset. The top three and six NetFlow features identified by the chi-square technique are sufficient for both DFF and RF classifiers to achieve maximum performance.

\end{multicols}
\begin{table}[ht]\scriptsize
\centering
\caption{CSE-CIC-IDS2018 and NF-CSE-CIC-IDS2018 full metrics}
\label{cicmetrics}
\begin{tabular}{|c|c|c|r|r|r|r|r|r|}
\hline
\textbf{Dataset} &
  \textbf{Classifier} &
  \textbf{\begin{tabular}[c]{@{}c@{}}Features\\ (Count)\end{tabular}} &
  \multicolumn{1}{c|}{\textbf{Accuracy}} &
  \multicolumn{1}{c|}{\textbf{AUC}} &
  \multicolumn{1}{c|}{\textbf{\begin{tabular}[c]{@{}c@{}}F1\\ Score\end{tabular}}} &
  \multicolumn{1}{c|}{\textbf{DR}} &
  \multicolumn{1}{c|}{\textbf{FAR}} &
  \multicolumn{1}{c|}{\textbf{\begin{tabular}[c]{@{}c@{}}Prediction\\ Time (\textmu s)\end{tabular}}} \\ \hline
\multirow{4}{*}{\textbf{CSE-CIC-IDS2018}}    & \multirow{2}{*}{\textbf{DFF}} & \textbf{Full (77)} & 96.45\% & 0.9501 & 0.87 & 81.31\% & 0.91\%  & 3.91 \\ \cline{3-9} 
                                             &                               & \textbf{CHI (6)}   & 92.45\% & 0.9461 & 0.79 & 86.31\% & 6.48\%  & 2.64  \\ \cline{2-9} 
                                             & \multirow{2}{*}{\textbf{RF}}  & 
                                  \textbf{Full (77)} & 98.01\% & 0.9668 & 0.93 & 94.79\% & 1.43\%  & 4.40 \\ \cline{3-9} 
                                             &                               & \textbf{CHI (3)}   & 98.36\% & 0.9654 & 0.94 & 93.95\% & 0.86\%  & 3.14  \\
                                              \hline
\multirow{4}{*}{\textbf{NF-CSE-CIC-IDS2018}} &

\multirow{2}{*}{\textbf{DFF}} 
& 
                                             \textbf{Full (8)}  & 85.74\% & 0.9256 & 0.61 & 92.74\% & 15.23\% & 4.82  \\ \cline{3-9}
                                             &                               & \textbf{CHI (3)}   & 85.61\% & 0.9155 & 0.61 & 93.25\% & 15.45\% & 4.40  \\
                                             
                                             \cline{2-9}
                                             & \multirow{2}{*}{\textbf{RF}}  & \textbf{Full (8)}  & 95.51\% & 0.9512 & 0.84 & 94.61\% & 4.36\%  & 8.17  \\ \cline{3-9} 
                                             &                               & \textbf{CHI (6)}   & 95.51\% & 0.9512 & 0.84 & 94.60\% & 4.36\%  & 7.62  \\ \hline
\end{tabular}
\end{table}
\begin{multicols}{2}
\subsection{Discussion}
\label{diss}

\begin{figure*}[ht]
\begin{subfigure}{.5\textwidth}
  \centering
  \includegraphics[width=8cm, height=4cm]{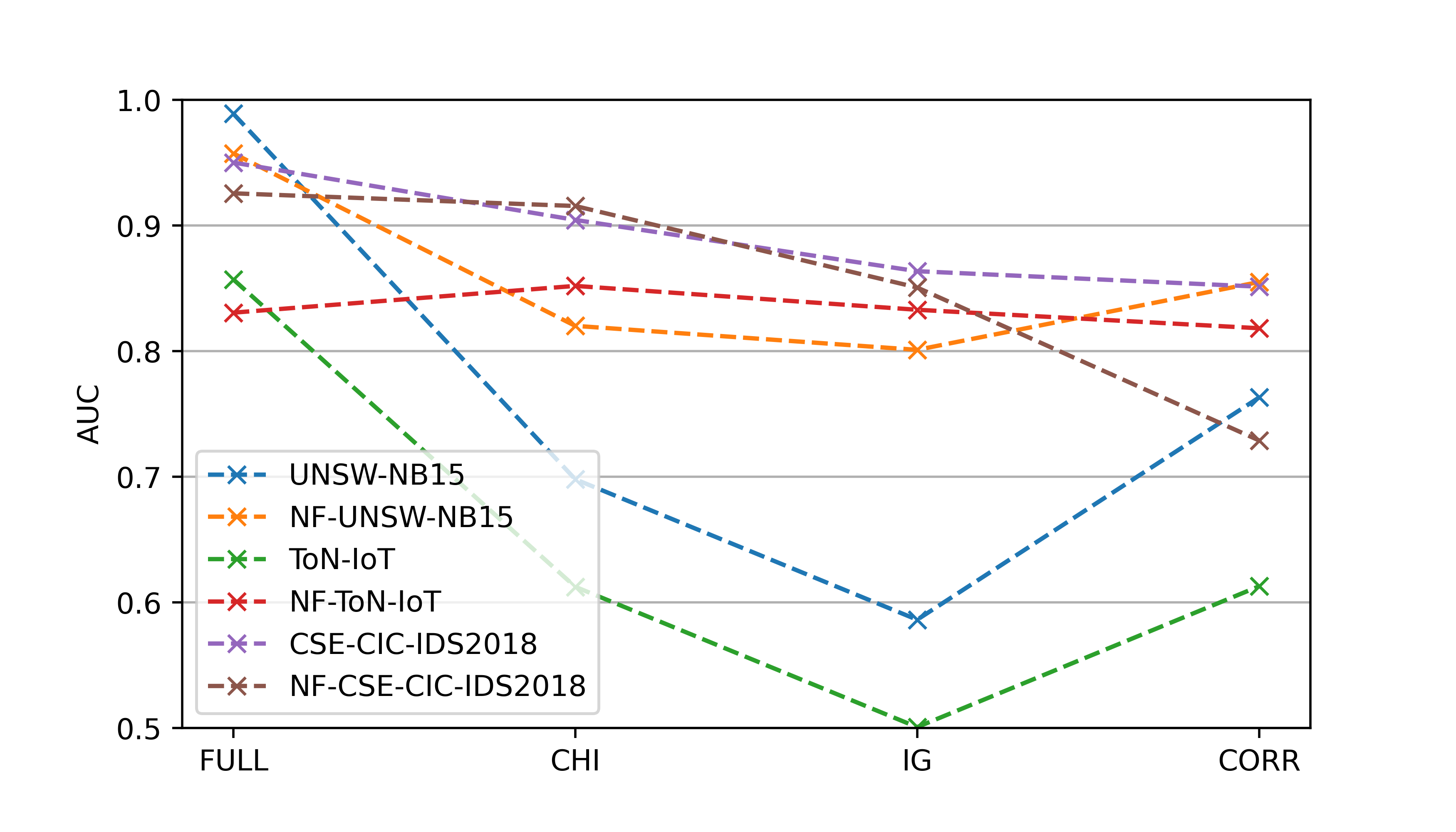}  
  \caption{Top three features using DFF}
  \label{fig:dnno}
\end{subfigure}
\hfill
\begin{subfigure}{.5\textwidth}
  \centering
  \includegraphics[width=8cm, height=4cm]{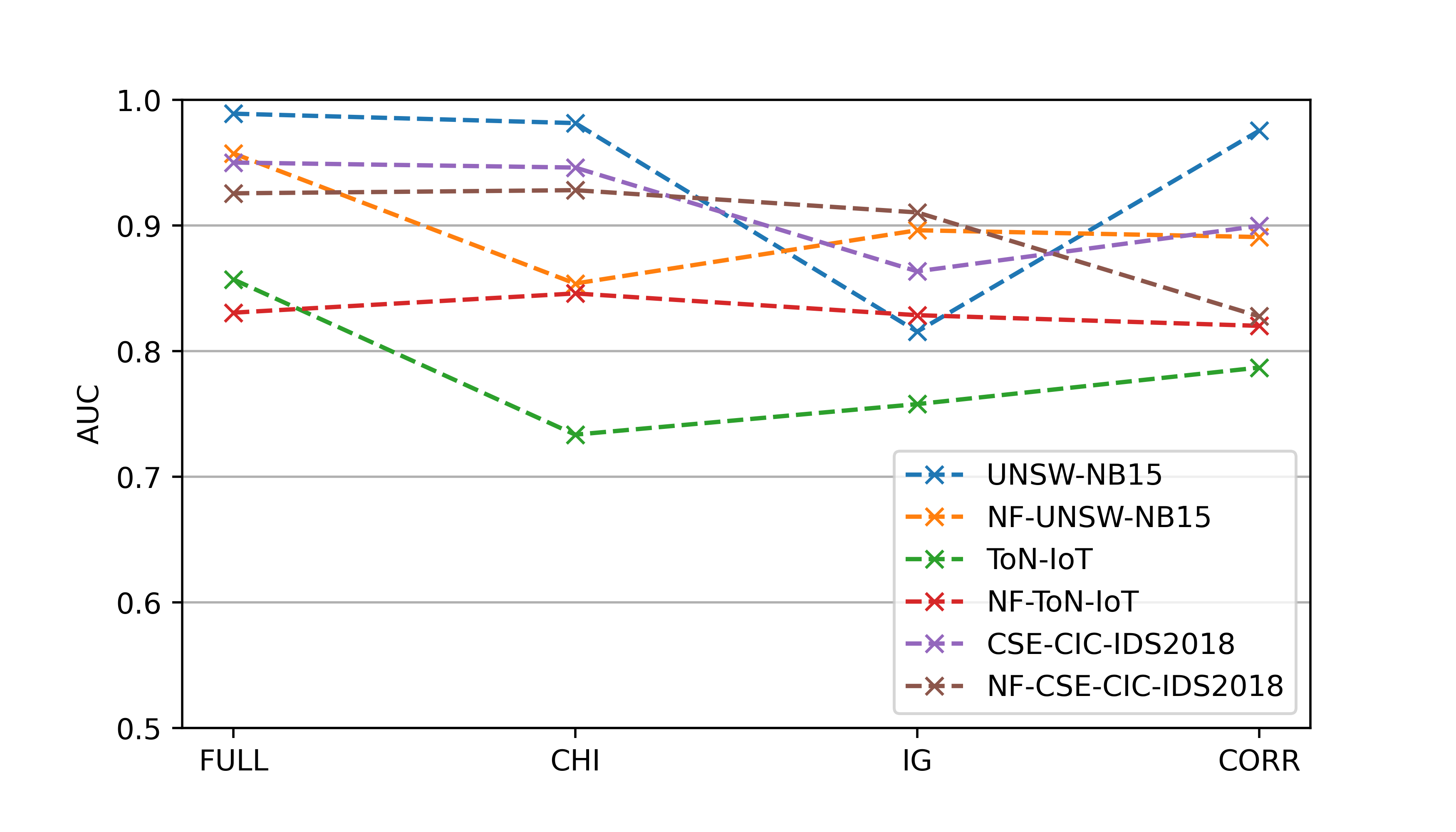}  
  \caption{Top six features using DFF}
  \label{fig:cnno}
\end{subfigure}
\hfill
\begin{subfigure}{.5\textwidth}
  \centering
  \includegraphics[width=8cm, height=4cm]{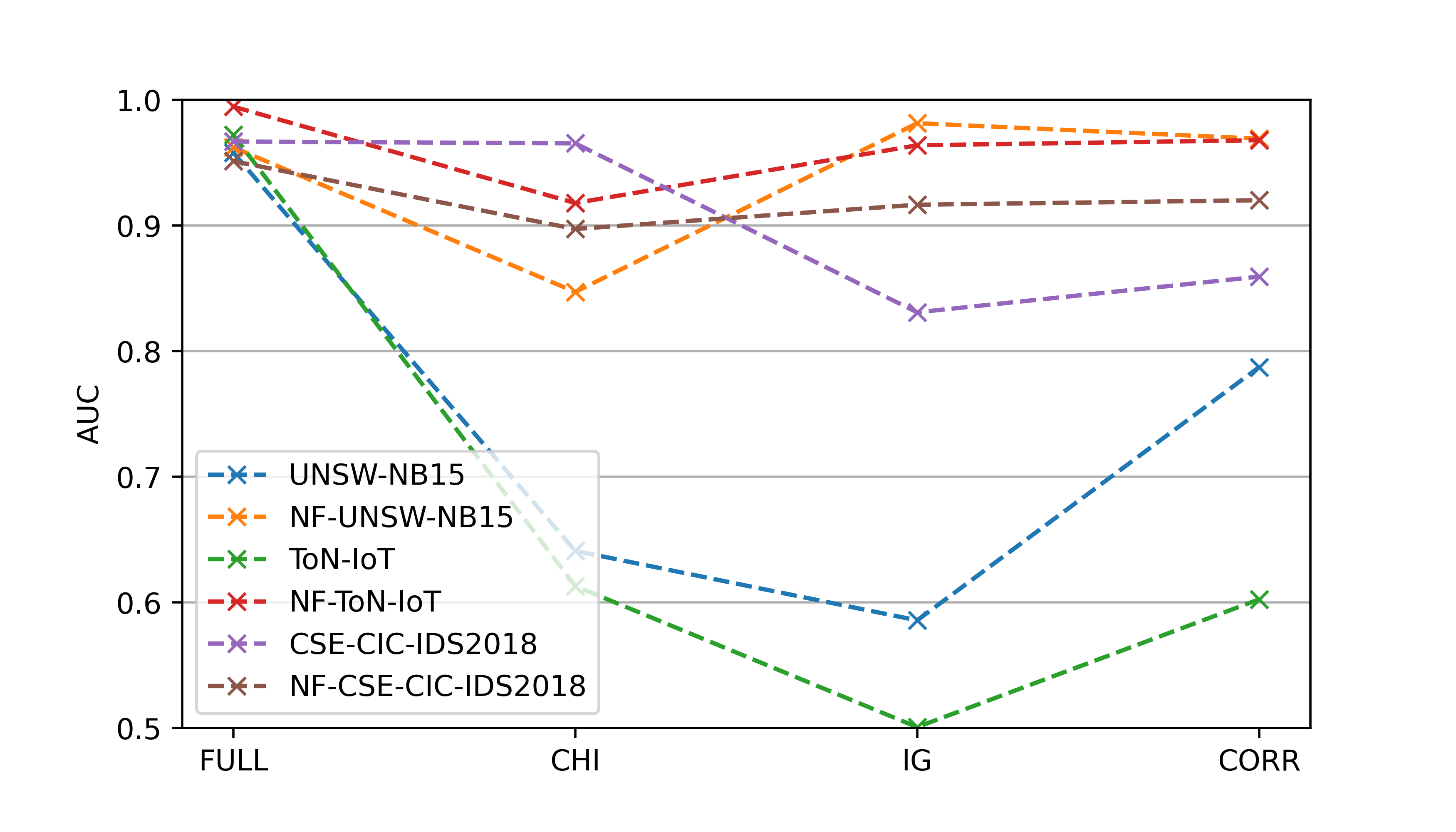}   
  \caption{Top three features using RF}
  \label{fig:dto}
\end{subfigure}
\hfill
\begin{subfigure}{.5\textwidth}
  \centering
  \includegraphics[width=8cm, height=4cm]{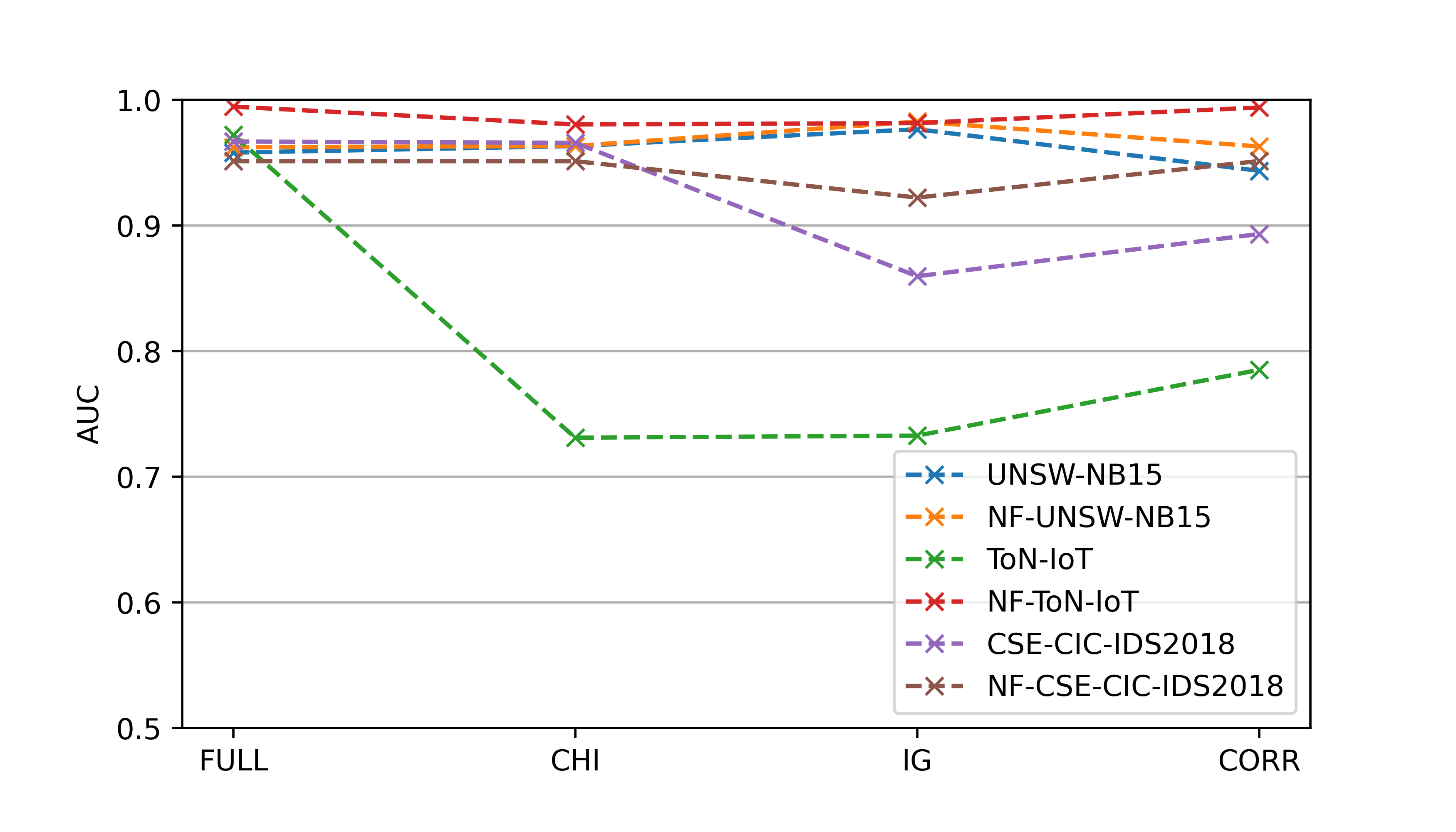} 
  \caption{Top six features using RF}
  \label{fig:nbo}
\end{subfigure}
\caption{performance of feature selection techniques across 6 NIDS datasets}
\label{summary}
\end{figure*}

Figure \ref{summary} provides a summary view of our results of the feature analysis. Figure (a) shows the AUC score achieved by the DFF classifier using the top three features based on the three feature selection algorithms. For comparison, the data point on the left shows the AUC score achieved with the full features. The results are shown for the six considered NIDS datasets. We observe significantly varying results for the different feature selection algorithms. For example, for the CSE-CIC-IDS2018 dataset, the top three features set chosen by the chi-square algorithm achieve an AUC score of 0.9654 using the RF classifier that is close to the one achieved with the entire set of features. In contrast, the top three features sets selected by the chi-square algorithm for the UNSW-NB15 and ToN-IoT datasets suffer a significant performance drop compared to the entire set of features. 

Figure \ref{summary} (c) shows the corresponding results for the RF classifier. Again, we observe that for some combinations of datasets and feature selection algorithms, e.g. (CSE-CIC-IDS2018, chi-square) and (NF-UNSW-NB15, IG), the top three feature subset can achieve the same AUC score as the complete feature set, or even higher in some cases. In contrast, for some combinations such as (ToN-IoT, IG) and (UNSW-NB15, IG), the top three features set achieve significantly reduced performance. Figures \ref{summary} (b) and (d) show the same results for the subsets of features consisting of the top six ranking features for the DFF and RF classifiers, respectively. We can observe that for most datasets, the top six features set can match the AUC score of the complete feature set if the best respective feature selection ranking is chosen. The only noticeable exception in this regard is the ToN-IoT dataset, where the top six feature suffers a significant drop in AUC score compared to the complete feature set, particularly for the RF classifier.

Our results show significant variability across the different considered datasets and feature selection algorithms to the network attack detection performance of varying-size feature subsets. The results demonstrate that there is no simple one-size-fits-all approach to feature selection and that careful analysis must be performed. Suppose the feature selection algorithm is carefully chosen and matched with the NIDS data set. In that case, a relatively minor feature subset can achieve the detection performance of the complete feature set. Such a significantly reduced feature set has the benefit of reduced resource consumption for feature extraction and storage and attack detection due to the reduced complexity of the classifier model and prediction times. More broadly, this is likely particularly relevant in IoT scenarios with resource-limited devices.

\section{Conclusion}
This paper has comprehensively evaluated feature importance across six NIDS datasets. Three feature selection techniques, i.e., chi-square, IG and correlation, have been utilised to rank the features in terms of predictive power and evaluated using DFF and RF classifiers, resulting in 414 experiments and data points. A key finding of these experiments is that a small subset of features can achieve the same or even higher detection performance than the complete set of features. This has the potential for a significant reduction in model complexity and computational and storage costs for feature extraction, which is critical in resource-constrained IoT environments. Another significant result of this paper is that there is no simple or general rule for choosing the optimal feature set and size. This is due to the high degree of variability in terms of feature importance across the different datasets and classifiers. 

A careful analysis has to be performed for the scenario considered to find the minimal feature set that can achieve the best classification performance. A final finding of our research is that a high degree of care must be taken when ML-based network intrusion detection algorithms and models are evaluated on synthetic NIDS datasets. As we have demonstrated, some features have unrealistically high predictive power, such as the TTL-based features in the UNSW-NB15 dataset, and should be removed before classification experiments. This is critical to obtain reliable evaluation results that can generalise to realistic network scenarios. This paper's findings further support our case for advocating a careful and extensive feature analysis when developing and evaluating ML-based NIDSs.

\end{multicols}
\section*{Ethics declarations}
\subsection*{Ethical approval}
This article does not contain any studies with human participants or animals performed by any authors.

\subsection*{Conflict of interest}
The authors have no competing interests to declare relevant to this article's content.

\subsection*{Funding declaration}
No funding was received to assist with the preparation of this manuscript.

\subsection*{Data availability}
All data generated or analysed during this study are included in this published article Sarhan, M., Layeghy, S., Moustafa, N. and Portmann, M., 2020. Netflow datasets for machine learning-based network intrusion detection systems. In Big Data Technologies and Applications (pp. 117-135). Springer, Cham.

\subsection*{Funding declaration}
No funding was received to assist with the preparation of this manuscript.

\subsection*{Author Contribution}
MS and SL wrote the main manuscript text. MP designed the framework and evaluation methodology. All authors reviewed the manuscript.

\bibliography{mybibfile}

\end{document}